\documentstyle[12pt]{article}
\def\be{\begin{equation}}
\def\ee{\end{equation}}
\def\bea{\begin{eqnarray}}
\def\eea{\end{eqnarray}}
\def\a{\alpha}
\def\b{\beta}
\def\l{\lambda}
\def\t{\tau}

\def\pa{\partial}
\def\e{\varepsilon}
\def\d{\delta}
 
\def\D{\Delta}
\def\L{\Lambda}

\begin{document}
 
\author{Hans-J\"urgen Schmidt}

\title{Fourth order gravity: equations, history, and applications to 
cosmology\footnote{This paper 
is a shortened version of  my (still unpublished)
``Lectures on Mathematical Cosmology'', see 
its preprint version gr-qc/0407095. Parts of my paper with
R. Schimming ``On the history of fourth order metric theories of 
gravitation''   NTM-Schriftenr. Gesch. Naturw., Tech., Med. 
(Leipzig) {\bf 27} (1990)  41,  gr-qc/0412038, are incorporated.}}

\date{42nd Karpacz Winter School, February 6 - 11, 2006}

\maketitle

\centerline{Universit\"at Potsdam, Institut f\"ur Mathematik, Am
Neuen Palais 10} 
 \centerline{D-14469~Potsdam, Germany,  E-mail:
 hjschmi@rz.uni-potsdam.de}
\centerline{http://www.physik.fu-berlin.de/\~{}hjschmi }

\begin{abstract} 
The field equations following from a Lagrangian $L(R)$
will be deduced and solved for special cases. If $L$ is a non-linear
function of the curvature scalar, then these equations are of fourth order
in the metric. In the introduction  we  present the history of these equations 
beginning with the paper of H. Weyl from 1918, who first discussed
them as alternative to Einstein's theory. In the third part, we give details 
about  the cosmic no hair theorem, i.e., 
the details how within fourth order gravity with $L= R + R^2$ the 
inflationary phase of cosmic evolution turns out to be a transient attractor. 
Finally, the Bicknell theorem, i.e. the conformal relation from fourth order
gravity to scalar-tensor theory, will be shortly presented.  
\end{abstract}

\section{To the history of fourth order gravity}

From the advent of the general relativity theory (GRT) in 1915 by 
Albert Einstein (1879-1955) until today numerous geometrized theories 
of gravitation have been proposed.
Here, we shall review the history of a class of theories which is 
conceptually rather close to GRT:

\noindent 
- The gravitational field is described by a space-time metric only.

\noindent 
- The field equation follows from a Hamiltonian principle. The 
Lagrangian $L$ is a quadratic scalar in die Riemannian curvature of 
the metric.

\noindent 
- The constants appearing in this ansatz are adjusted 
such that the theory is compatible with experimentally established facts.
Hence, the Lagrange function reads
\be
L = aR^2 + b R_{ij}R^{ij} + kR + \Lambda
\ee
with constants $a, \  b, \  k, \  \Lambda$ where $a$ and $b$ do 
not vanish simultaneously.  The variational derivative of  
$R_{ijkl}  \,  R^{ijkl}  $  with respect to the metric 
can be linearly  expressed by the variational derivatives of   
$R_{ij} \, R^{ij}  $ and of  $R^2$  [1]. Thus we may omit  $R_{ijkl}  \,  R^{ijkl}  $
 in (1.1) without loss of  generality. The theory is scale-invariant if and
 only if   $ \Lambda \cdot k = 0$.  It is even conformally invariant if and only if 
$\Lambda  = k = 0$ and $3a +b = 0$. 
The field equation following from $L$ eq. (1.1) is of fourth order, 
i.e.   it contains derivatives up to the fourth order of the components 
of the metric with respect to the space-time coordinates. 

The fourth order metric theories of gravitation are 
a very natural modification of the GRT. Historically, they 
have been introduced as a specialization of Hermann Weyl's 
(1885-1955) nonintegrable relativity theory from 1918 [2].  
Later on, just the fourth order theories became
interesting and more and more physical motivations supported 
them: The fourth order terms can prevent the big bang singularity 
of GRT; the gravitational potential of a point mass is 
bounded in the linearized case; the inflationary cosmological 
model is a natural outcome of this theory. But all the 
arguments from classical physics were not so convincing as 
those from quantum physics: the quantization of matter 
fields with unquantized gravity background leads to a 
gravitational Lagrangian of the above form [3]. Moreover, 
fourth  order theories turned out to be renormalizable at the 
one-loop quantum level [4], but at the price of losing the 
unitarity of the S-matrix.  These circumstances caused a boom of fourth
 order gravity (classical as well as quantum) in the seventies. We will 
stop our record of the history before this boom. We restrict ourselves to 
the purely metrical theories (i.e., the affinity is always presumed 
to be Levi--Civita) and want only to mention here that fourth
 order field equations following from a variational principle
 can be formulated in scalar-tensor theories, theories with 
independent affinity, and other theories alternative to GRT
 as well.

\subsection{Papers inspired by Weyl's theory}

In 1918, soon after Albert Einstein's proclamation of the GRT,
 Hermann Weyl proposed a new kind of geometry and a
 unified theory of gravitation and electromagnetism based on
 it. He dwelled on the matter in a series of papers [2, 5-10] 
until it became superseded by the modern gauge field interpretation
 of electromagnetism [11-13]. Note that the gauge concept together
 with the words ``Eichung" (gauge) and ``Eichinvarianz" (gauge invariance) 
came into use in theoretical physics through Weyl's ansatz. For a 
broader discussion and evaluation we refer to [14]. A. Einstein [15] pointed 
out that the nonintegrability of the lengths of vectors 
under Weyl-like parallel propagation contradicts to physical 
experience. His argument has been refuted not earlier than 
in 1973: P.  Dirac  discusses 
the possibility of a varying gravitational constant. He writes:

``Such a variation would force one to modify  
Einstein's theory of gravitation. It is proposed that the 
modification should consist in the revival of Weyl's geometry, 
in which lengths are nonintegrable when carried around 
closed loops, the lack of integrability being connected 
with the electromagnetic field".  [16, p. 403]

H. Weyl's aesthetically very appealing modification of 
 GRT unfortunately does not directly
describe the real dynamics of fields and particles; 
however it deeply influenced the ``dynamics
of theories". By this we mean that various fundamental ideas 
have been formed or promoted by Weyl's papers:

\noindent 
- the search for alternatives to the GRT based on geometrization;

\noindent 
- the unification of the interactions or forces 
of nature, beginning with gravity and electromagnetism;

\noindent 
- field theories based on the geometry of an affine connection;

\noindent 
- conformal geometry and conformally invariant field theories;

\noindent 
- the gauge field idea, and 

\noindent 
- fourth order gravitational field equations.

Here we are interested just in the last item. Weyl required 
the Lagrangian to be a polynomial function of the curvature 
and to be conformally invariant. He states:
 ``Dies hat zur Folge, dass unsere Theorie wohl 
auf die Maxwellschen elektromagnetischen nicht 
aber auf die Einsteinschen Gravitationsgleichungen 
f\"uhrt; an ihre Stelle treten Differentialgleichungen 4. 
Ordnung." [2, p. 477] 

The ambiguity in the concrete choice of  $L$ appeared as a 
difficulty which is opposed to the spirit of 
unification: any linear combination of  $R^2$ and  $R_{ij} \, R^{ij}  $
  would do. The variation of  $R_{ij} \, R^{ij}  $  or of
$R_{ijkl}  \,  R^{ijkl}  $  with respect to the vector field yields 
Maxwell-like equations, 
while for the choice of  $R^2$ an electromagnetic Lagrangian 
$F_{ij} \, F^{ij}  $
together with  a coupling constant $\alpha$ has to be added by 
hand:   $L = R^2 + \alpha  F_{ij} \, F^{ij}  $  [6, 2]. Weyl himself 
favoured different Lagrangians in different papers. 
Moreover,  he took trouble to produce results compatible 
with Einstein's GRT.  For this aim he destroyed the 
conformal invariance by a special 
gauge. Ernst Reichenb\"acher criticizes:
 ``Um so auffallender ist es, dass Weyl in dem von ihm
durchgerechneten Beispiel f\"ur die Wirkungsfunktion durch 
Festlegung der Eichung vor der Variation den Grundsatz 
der  Eichinvarianz durchbricht." [17, p. 157].
A more detailed analysis of the theory was necessary then. 
R.  Weitzenb\"ock [18] produced and studied all scalar invariants 
of the curvature in Weyl's geometry. W.  Pauli 
 [19, 20] and a little later F.  J\"uttner  [21] calculated the spherically symmetric 
static  gravitational field for variants of Weyl's theory. 

Pauli [20, S. 748]  comes to an important conclusion:
``Hiernach ist klar, dass aus Beobachtungen der 
Merkurperihelbewegung und der Strahlenablen\-kung, die
mit Einsteins Feldgleichungen im Einklang sind, niemals 
ein Argument gegen Weyls Theorie entnommen werden kann, 
wenigstens solange die letztere eine der
 drei Invarianten    $R_{ij} \, R^{ij}  $, $R^2$, 
$R_{ijkl}  \,  R^{ijkl}  $   
 als Weltfunktion zugrunde legt." 

In other words, fourth order gravitational field equations 
following from (1.1) are not falsifiable by experimental physics! 
Pauli [20, 22] and other authors  did not even consider the vector 
field (i.e. assumed it to be equal to zero) in Weyl's theory, 
thus  making it unaffected by the criticism of nonintegrability [15]. 
R. Bach [23] realized the possibility to keep the conformal invariance 
in a purely metrical theory: a Lagrangian 
$ L = C_{ijkl}  \,  C^{ijkl}  $   or   equivalently,
$ L  = R^2 - 3R_{ij} \,  R^{ij} $
yields a conformally invariant field equation for the metric, later 
on called ``Bach's equation". In a similar spirit and in the 
same year 1921, Albert Einstein [24] proposed a 
conformally invariant theory. His expressions suffer from being 
non-rational in the metric. This theory is sometimes cited but 
has never been studied in details.

Reichenb\"acher [25, 17] proposed a variant of Weyl's theory based 
on a non-rational Lagrangian resembling nonlinear 
Born--Infeld electrodynamics.

In [26], also $L = R^2$ is used to get a field-theoretical 
model for the electron, but the fourth order terms are
 lost by an error in the calculations.

C. Lanczos  [27] tried a 
programme of ``Electromagnetismus als nat\"urliche 
Eigenschaft der Riemannschen Geometrie". 
He also assumed the vector field in Weyl's theory to be zero, but 
reintroduced 
it then in an alternative way as a set of Lagrangian multipliers. 
Unfortunately, Lanczos was, working 
with hyperbolic differential equations, misled by a formal analogy with
 elliptic differential equations. He varied the speculations with
 Lagrangian multipliers in a series of papers [28-34]. 
To take it positive, many useful mathematical formulas for fourth order 
theories 
resulted from Lanczos' work. Particularly, the paper [1] became 
a  ``citation classic".

In the twenties, the programme of classical field theory 
with its two cornerstones geometrization and unification lost 
some of its attractiveness in virtue of the quickly 
progressing quantum theory, cf. [35]. Moreover, there were the 
refutation of Weyl's theory and objections to fourth 
order equations. Lanczos expressed them as follows:

``Der Grund, weshalb diese Untersuchungen nicht weiter gediehen 
sind und zu keinem Fortschritt f\"uhrten, lag an zwei Momenten. 
Einerseits war es entmutigend, dass  man zumindest drei 
anscheinend gleichwertige Invarianten zur Verf\"ugung
 hatte: $R^2$, $R_ {\alpha  \beta}R^{\alpha \beta}$,
$R_ {\alpha \beta \gamma  \delta }R^{\alpha \beta  \gamma  \delta}$,
 ohne ein plausibles Auswahlprinzip zwischen
ihnen zu besitzen. Andererseits erscheinen diese Gleichungen, 
solange man ihre innere Struktur nicht verstehe, als 
Differentialgleichungen vierter Ordnung f\"ur  die die $g_{ik}$
von einer Kompliziertheit sind, die f\"ur jede weitere Schlussfolgerung 
ungeeignet ist." [27, p.75]

Similarly, Bergmann argues in his text-book [36] that, 
first, fourth order equations admit too many solutions  
and, second, their Lagrangian is rather ambiguous.
This situation explains why only  few papers on fourth 
order gravitation appeared in the period from the thirties to 
the sixties and why these did not follow the actual trends at 
their time. H. A. Buchdahl dealt with the subject in the period 
1948-1980. In his papers he covered the following problems:

\noindent 
- Invariant-theoretical considerations continuing those 
of Weitzenb\"ock and Lanczos [37];

\noindent 
- General expressions for the variational derivatives of 
Lagrangians built from the curvature and, possibly, its 
derivatives are obtained [38-42].

\noindent 
- Einstein spaces are solutions of a rather general class of 
fourth order equations [43, 44];

\noindent 
- Static gravitational fields in fourth order theories [45];

\noindent 
- Cosmological solutions in theories where the Lagrangian is 
a function of  the scalar curvature [46, 47];

\noindent 
- Conformal gravity [48];

\noindent 
- Reinterpretation of some fourth order equations in 
five dimensions [49].

 A.  S. Eddington  in 1921, see [50], 
and E. Schr\"odinger   in 1948, see [51], 
also discussed gravitational field equations of fourth 
order to get field theoretical particle models, i.e., they 
tried to realize Einstein's particle programme.

\subsection{A new view}

Fourth order metric theories of gravitation have been 
discussed from 1918 up to now. One original motivation 
was the scale invariance of the action, a property which 
does not hold in GRT. Another motivation was the search 
for a unification of gravity with electromagnetism, which 
is only partially achieved with the Einstein-Maxwell system. 
There was no experimental fact contradicting GRT which could 
give motives for replacing it by a more complicated theory.
But a lot of problems appeared:
 
\noindent 
1. The Lagrangian became ambiguous in sharp contrast to 
the required unification.

\noindent 
2. The higher order of the field equation brought
 mathematical problems (2.1.) in the search for solutions and
 physical problems (2.2.) for the interpretation of the additional 
degrees of freedom.

\noindent 
3. The well-founded Newtonian theory of gravitation did not 
result as the weak-field limit of
scale invariant fourth order gravity.

The third problem was the last of these to be realized but the 
first to be solved, both in 1947: One has to break the scale invariance
 of the theory by adding the Einstein--Hilbert action to the 
purely quadratic Lagrangian. Then, up to an exponentially 
small term, the correct Newtonian limit appears [52].

The original scale invariant theory then, again emerges as 
the high-energy limit of that sum. The items 1., 2. and 
the absence of experimental facts contradicting GRT seemed 
to restrain die research on these theories already in the twenties. 
Only in 1966 a renewed interest in these theories arose 
in connection with a semiclassical description of quantum 
gravity [53-55]. The coefficient of the quadratic term
 became calculable by a renormalization procedure, thus 
solving problem 1, at least concerning the vacuum equation. 
Further, the fact that fourth order gravity is one-loop 
renormalizable in contrast to GRT; a fact which was 
realized in 1977, [4] initiated a boom of research. 
It is interesting to observe that it is just the scale invariance 
of the curvature squared terms -- the original  motivation -- which 
is the reason for the renormalizability. Also the 
latest fundamental theory -- the superstring theory -- gives 
in the field theoretical limit (besides other terms) just 
a curvature-squared contribution to the action [56, 57]. The 
use of modern mathematics and computers has led 
to a lot of results to clarify the structure of the space 
of solutions thus solving problem 2.1. in the eighties. The 
more profound problem 2.2 has now three kinds of answers:

a) In spite of the higher order of the differential equation, a 
prescribed matter distribution plus the $O (1/r)$-behaviour of 
the gravitational potential suffice -- such as it takes place 
in Newtonian theory -- to determine the gravitational potential 
for isolated bodies in a unique way for the weak-field 
slow-motion limit, [52, 53]

b) the observation that the additional degrees of freedom are 
just the phases of damped oscillations which become 
undetectably small during the cosmic evolution, and, by the 
way, can solve the missing mass problem and 
prevent the singularity problem of GRT [58], and

c) it is supposed that there exist massive gravitons besides the 
usual massless gravitons known from GRT, but they 
are very weakly coupled [59].

The last point to be mentioned is the experimental testability: 
In the recent three years many
efforts have been made to increase the accuracy in determining 
the constants $G$, $\alpha$ and $l$ if the gravitational potential 
is assumed  (also in other theories than fourth 
order gravity) to be
$ Gm r^{-1} (1 + \alpha e^{-r/l} ) \, . $
The term proportional to  $\alpha$, the ``fifth force", can 
be interpreted as the fourth order correction to GRT. Up to 
now, it has not been possible to exclude  $\alpha =0$ by 
experiments [60-62]. 

 In 1990, we concluded our paper on the history of fourth order 
gravity by  saying: Fourth order gravity theories will 
remain an essential link between GRT and quantum gravity 
for a long time. Now, 16 years later, one can prove the correctness  of this
 prediction by looking at references [63-99], [102], [105-109]
 which are all deal with 
 a variant of fourth order gravity or its equivalents  and its
 cosmological consequences and which is still far from being  complete. 

\medskip

\section{Scalar fields and $f(R)$  cosmology}
\setcounter{equation}{0}

Following the paper ``Comparing selfinteracting scalar fields and $R +
R^3$   cosmological models'',   gr-qc/0106035, see [82], 
 we generalize the well-known analogies between $m^2 \phi^2$ 
 and $R + R^2$  theories to include   the self-interaction 
$ \l \phi^4$-term for the scalar field. It turns out to be the $R + R^3$
 Lagrangian which gives an appropriate model for it. Considering a spatially 
flat Friedmann cosmological model, common and different properties 
of  these models are discussed, e.g., by linearizing around 
a ground state the masses of the corresponding  spin 0-parts coincide. 
 Then  we prove   a general conformal equivalence
 theorem between a Lagrangian
$ L = L(R)$, $ L'L'' \ne  0$, and a minimally coupled scalar
 field in a general potential.
 This theorem was independently deduced 
 by several persons, and it is now known as Bicknell theorem [68].

\subsection{Introduction to scalar fields}

For the gravitational Lagrangian
\be\label{7.1}
     L = (R/2 + \b R^2)/8\pi G \,  ,    
\ee
where $\b$ is some free but constant  parameter, the value  
\be\label{7.1f}
R=R_{\rm  crit} = - 1/4\b 
\ee 
 is the critical value of the curvature scalar. It is  defined by 
\be\label{7.1a}
\pa L/ \pa R = 0 \, .
\ee 
 In regions where 
\be
R/ R_{\rm  crit} < 1
\ee
 holds,  we can define 
\be
 \psi  = \ln (1 - R/R_{\rm  crit})
\ee
  and 
\be\label{7.1d}
 \tilde g_{ij}   =   \left(  1 - R/R_{\rm  crit} \right)   g_{ij} \, .
\ee
 In units where  $8\pi G = 1$ we now obtain from the Lagrangian
 eq. (\ref{7.1}) via the conformal transformation eq. (\ref{7.1d})
 the transformed Lagrangian 
\be\label{7.2}
 \tilde   L = \tilde R/2  - 3 \tilde g^{ij} \psi_{;i} \psi_{;j}/4 - \left(1
 - e^{-\psi}\right)^2 / 16 \b
\ee
being equivalent to $L$,   cf.  Whitt [85].

For $\b < 0$, i.e., the absence of 
tachyons in $L$ eq. (\ref{7.1}), we have massive  gravitons of mass 
\be\label{7.2a}
m_0 = (-12 \b)^{-1/2}
\ee
 in $L$, cf. Stelle [4]. For the weak field limit, the potential
 in  eq. (\ref{7.2}) can be simplified to be $\psi^2/ (16\cdot \b )$,
 i.e., we have got a minimally coupled scalar field whose mass 
is also $m_0$.  The superfluous factor 
3/2 in  eq. (\ref{7.2}) can be absorbed by a redefinition of $\psi$.
Therefore, it is not astonishing, that all results concerning 
the weak field limit for both $R + R^2$-gravity without 
tachyons and Einstein gravity with a minimally coupled 
massive scalar field exactly coincide. Of course, one cannot 
expect this coincidence to hold for the non-linear region, too, 
but it is interesting to observe which properties hold there also.

We give only one example here: we consider a cosmological 
model of the spatially flat Friedmann type, start integrating 
at the quantum boundary, which is obtained by  
$ R_{ijkl}R^{ijkl} $   on the one hand, and $T_{00}$
 on the other hand, to have Planckian values, 
 with uniformly distributed initial conditions and look 
whether or not an inflationary phase of 
the expansion appears. In both  cases we get the following result: The probability $p$
 to have sufficient inflation  is about $p = 1 - \sqrt{\l} m_0/m_{\rm Pl}$,  i.e., 
$p = 99.992 \% $ if we take $m_0 = 10^{-5} m_{\rm Pl}$ 
 from GUT and $\l = 64$, where $e^\l$ is the linear multiplication factor of 
inflation.

From Quantum field theory, however, instead of the massive 
scalar field, a Higgs field with self-interaction turns out to 
he a better candidate for describing effects of  
the early universe. One of the advances of 
the latter is its possibility to describe a 
spontaneous breakdown of symmetry. In the following, we 
try to look for a purely geometric model for this Higgs field which is analogous to the above 
mentioned type where $L = R + R^2$  modelled a massive scalar field.

\subsection{The Higgs field}\label{s72}

For the massive scalar Field $\phi $ we have the mater Lagrangian 
\be\label{7.3a}
L_m = - \left(  \phi_{;i} \phi^{;i} - m^2 \phi^2
\right) /2 \, , 
\ee
and for the Higgs field to be discussed now,
\be\label{7.3b}
     L_\l = - \left(
 \phi_{;i} \phi^{;i} + \mu^2 \phi^2 - \l \phi^4/12 
\right) /2 \, . 
\ee
The ground states are defined by $\phi =$ const. and
$\pa L/\pa \phi =0$. This means  $\phi = 0$  for the scalar field, and 
the three ground states $\phi = \phi_0 = 0$, and
\be\label{7.4a}
\phi = \phi_\pm = \pm \sqrt{6 \mu^2/\l}
\ee
  for the Higgs field.

The expression 
\be\label{7.3}
     \left(  \pa^2 L/ \pa \phi^2 \right)^{1/2}
\ee
represents  the effective mass at these points. This gives the
 value $m$ for the scalar field  eq. (\ref{7.3a}), so justifying the notation. 
 Further, eq. (\ref{7.3}) give mass  $\, i \, \mu$  at $\phi = 0$  and 
 $\sqrt{2} \mu $  at   $\phi = \phi_\pm$ 
 for the Higgs field  eq. (\ref{7.3b}). The imaginary value of the mass 
at the ground state $\phi = 0$ shows the instability met there, and in 
the particle picture,  this gives rise to a tachyon.

To have a vanishing 
Lagrangian at the ground state     $ \phi_\pm$ eq. (\ref{7.4a})
we add  a   constant
\be\label{7.4}
\L  = -3 \mu^4/2 \l 
\ee
to the Lagrangian  eq. (\ref{7.3b}).  The final Lagrangian reads
\be\label{7.5}
L = R/2 + L_\l + \L 
\ee
 with $L_\l$  eq. (\ref{7.3b})  and $\L$  eq. (\ref{7.4}).

\subsection{The non-linear gravitational Lagrangian}

Preliminarily  we direct the attention to the 
following fact: on the one hand, for Lagrangians   (\ref{7.3a}),   (\ref{7.3b})
 and  (\ref{7.5})  the transformation $\phi \to - \phi$
 is a pure  gauge transformation, it does not change any 
invariant or geometric objects. On the other hand,
\be\label{7.6}
 R_{ijkl} \to -  R_{ijkl}
\ee
or simpler
\be\label{7.7}
R \to  -R
\ee
is a gauge transformation at the linearized level only: taking
\be
g_{ik} = \eta_{ik} +  \e h_{ik} \,   ,
\ee
where
\be
\eta_{ik} = {\rm diag} (1, -1, -  1, - 1) \, , 
\ee
then $\e \to - \e $ implies curvature inversion  eq. (\ref{7.6}). 
 To be strict: this is valid   at the linearized level in $\e$ only.
 On the other hand, the curvature inversion eq. (\ref{7.6}), 
 and  even its simpler version eq. (\ref{7.7}), fails to  hold quadratic  in $\e$. 
This corresponds to  fact that the $\e^2$-term in   eq. (\ref{7.2}), which   
corresponds  to the $\psi^3$-term in the development
 of $\tilde L$ in powers of $\psi$,  
  is the first one to break  the $\psi \to - \psi$ symmetry in  eq. (\ref{7.2}).

In fact, the potential is essentially 
\be
 \left( 1 - e^{( - x)} \right)^{2} \, .
\ee
Calculating this, the result reads 
\begin{eqnarray}
x^{2} - x^{3} + {\displaystyle \frac {7}{12}} \,x^{4} + \dots
\end{eqnarray}

Now, let us introduce the general non-linear Lagrangian $L = L(R)$  which 
we at the moment  only assume to be an analytical function of $R$.
 The  ground states are defined by $R =$ const., i.e.,
\be\label{7.8}
     L'R_{ik}- g_{ik} L/2   = 0    \, .
\ee
Here, $L' = \pa L/ \pa R$.

\subsection{Calculation of the ground states}\label{t731}

From  eq. (\ref{7.8}) one immediately sees that $\pa L/ \pa R =0 $ 
defines critical values of the curvature scalar. 
For these values $ R = R_{\rm  crit}$    it holds: 
For $ L(R_{\rm  crit})  \ne  0 $  no such ground state exists,  and for 
$ L(R_{\rm  crit}) =  0 $, we have  only one equation 
 $ R = R_{\rm  crit}$  to be solved with  10 arbitrary functions $g_{ik}$. 
 We call these ground states degenerated ones.  For $L = R^2$, 
 $ R_{\rm  crit}  = 0$, this has been discussed by Buchdahl  [41]. 
 Now, let us concentrate on the case $\pa L/ \pa R  \ne  0$. 
 Then $R_{ij} $  is proportional to $g_{ij}$  with a constant 
proportionality factor, i.e., each ground 
state is an Einstein space
\be\label{7.9}
R_{ij}= R \, g_{ij} /4 \,    , 
\ee    
 with a prescribed constant value $R$. 
Inserting  eq. (\ref{7.9}) into  eq. (\ref{7.8})  we get as condition for ground states
\be\label{7.11a}
            RL' = 2L \, .
\ee
     As an example, let $L$  be a third order polynomial
\be\label{7.10}
            L = \L  + R/2 + \b R^2 + \l R^3/12 \,  . 
\ee
We consider only Lagrangians with a positive 
linear term as we wish to reestablish Einstein gravity in 
the $\L \to   0$  weak field limit, and $\b  <  0$  to exclude tachyons there.

We now solve eq. (\ref{7.11a}) for the Lagrangian eq. (\ref{7.10}).
For $\l = 0$ we have,  independently of $\b$,  the only 
ground state $R = -4\L$. It is a degenerated one 
if and only if   $\b \L = 1/16$. That implies that for 
usual $R + R^2$ gravity eq. (\ref{7.1}), i.e. 
$\l = \L = 0$, we get  \, $ R = 0 $ \, as the 
only ground state;  it is a non-degenerated one.

Now, let $\l \ne 0$ and $\L =  0$. To get non-trivial 
ground states we need the additional assumption  
 $\l  > 0$. Then, besides $R = 0$, the ground states are
\be\label{7.11}
     R=R_\pm  = \pm \sqrt{6/\l}
\ee
being quite analogous to the ground states eq. (\ref{7.4a}) of the Higgs 
field  eq. (\ref{7.3b}). The ground state $R = 0$  is  not 
degenerated. Of course, this statement  is 
independent of $\l$  and holds true, as one knows, 
for $\l = 0$.  To exclude tachyons, we require $\b < 0$, then $R_-$  is not degenerated and 
$R_+$  is degenerated if and only if   $\b  = -\sqrt{6/\l}$.
 The case $\l \L  \ne  0$  will not be considered here.

\subsection{The field equation}

For $L = L(R)$  the variation
\be
\d \left( L \,  \sqrt{-g} \,  \right) / \d g^{ij} =0
\ee
 gives with $L' = \pa L/ \pa R $ the following fourth order gravitational
 field equation 
\be\label{7.14}
L' R_{ij} - g_{ij} L/2 + g_{ij } \Box L' - L'_{;ij}
=0                  \,  .
\ee
 It holds 
\be\label{7.15}
L'_{;ij} = L'' R_{;ij} + L'''\,  R_{;i} R_{;j} \, .
\ee
With  eq. (\ref{7.15}), the trace of  eq. (\ref{7.14})  reads
\be\label{7.16}
     L'R - 2L + 3L'' \Box  R + 3L'''\, R_{;k}R^{;k} = 0 \,   , 
\ee
i.e., with $L$   eq. (\ref{7.10}) 
\be
-2 \L  -R/2 + \l R^3/12 + 6 \b \Box R
 + \frac{3 \l}{2}  (R \Box R + R_{;k}R^{;k} )  =0 \,  .  
\ee
Inserting eqs.  (\ref{7.12}),  (\ref{7.13}) and  (\ref{7.15})
into the $00$-component of   eq. (\ref{7.14})  we get the equation
\bea\label{7.17}
0    = h^2/2 - \L /6 - 6 \b (2h \ddot h - \dot  h^2 + 
6h^2 \dot h ) \nonumber \\  + 3 \l (\dot h  + 2h^2)
 (6h \ddot h  + 19h^2 \dot h - 2\dot h^2 - 2h^4) \,  . 
\eea
The remaining components are a consequence of this one.

\subsection{The Friedmann   model}

Now we take as Lagrangian eq.   (\ref{7.10}) and as line element
\be\label{7.12}
 ds^2= dt^2 - a^2(t) (dx^2 + dy^2 + dz^2)   \, .
\ee
The dot denotes $d/dt$  and $h = \dot a/a$ and we have
\be\label{7.13}
          R =  -6 \dot h - 12h^2 \,   .
\ee
Here we only consider the spatially flat 
Friedmann  model  eq. (\ref{7.12}). Therefore, we can 
discuss only de Sitter  stages with $R < 0$, especially  the ground state 
 $R_+$  eq. (\ref{7.11}) representing an anti-de Sitter spacetime 
  does not enter our discussion, but $R_-$ does.

Now, let $\L =  0$. Solutions of   eq. (\ref{7.17})  with 
constant values $h$ are $h=  0$ representing 
flat spacetime  and in the case that $\l > 0$ also 
\be
h = \frac{1}{\sqrt[4]{24 \l} }
\ee
representing the de Sitter spacetime. These are  the non-degenerated ground states $R = 0$ 
and $ R = R_-   = - \sqrt 6/\l$,  respectively. Eq.   (\ref{7.17})  can be written as
\bea\label{7.20}
     0 = h^2(1 - 24 \l h^4)/2 + h \ddot h \left(
 1/m_0^2 + 18 \l (\dot h + 2 h^2) \right) \nonumber \\
 - 6 \l \dot h^3 + \dot h^2 (45 \l h^2 - 1/2m_0^2 )
+ 3 h^2 \dot h (1/m_0^2 + 36 \l h^2 ) \, .
\eea
First, let us consider the singular curve 
defined by the vanishing of the coefficient of   $\ddot h$  in  eq. (\ref{7.20})  in 
the $h - \dot h$-phase plane. It is, besides $h = 0$, the curve
\be\label{7.21}
\dot h =    -2h^2 - 1/18\l m_0^2
\ee
 i.e., just the curve 
\be
R = 1/3\l m_0^2 = -4 \b /\l
\ee
 which is  defined by $L'' = 0$, cf.   eq. (\ref{7.16}). This value coincides with 
 $R_+ $ if $\b  = - \sqrt{3 \l /8}$,  this value we need  not discuss here. Points 
of the curve  eq. (\ref{7.21})  fulfil  eq. (\ref{7.20})  for
\be
h = \pm \, 1 \, / \, 18\l m_0^3 \sqrt 3 \,  \sqrt{1-1/18\l m_0^4} 
\ee
only, which is not real because of $\l \ll m_0^4$. 
Therefore, the space of solutions is composed of at least two  connected 
components.

Second, for $h = 0$ we have $\dot h = 0$ or
\be\label{7.22}
\dot h = - 1/12 \l m^2   \, .
\ee
From the field equation we get: $h = \dot h =0$
 implies $h \ddot h \ge 0$,  i.e. $h$ does not 
change its sign. In a neighbourhood of  eq. (\ref{7.22}) 
 we can make the ansatz
\be
h = -t/12 \l m_0^2 + \sum_{n=2}^\infty \, a_n \, t^n
\ee
which has solutions with  arbitrary values $a_2$. This  means:  one can 
change from expansion to subsequent recontraction,  
but only through the ``eye of a needle"  eq. (\ref{7.22}). On the other hand, 
a local minimum of the scale factor 
never appears. Further,  eq. (\ref{7.22}) does not 
belong to the connected component of  flat spacetime.

But we are especially interested in the latter one, and therefore, we restrict to the subset 
$\dot h > \dot h$( eq. (\ref{7.21})) and need 
only to discuss expanding solutions $h \ge 0$. Inserting $\dot h = 0$,
\be
\ddot h =    h(24 \l h^4 - l)/(2/m_0^2 + 72\l h^2)
\ee
turns out, i.e., $\ddot h > 0$ for $h > 1/ \sqrt[4]{24 \l }$  only. 
All other points in the $h - \dot h$ phase plane are regular ones, and one can 
write $d \dot h / dh \equiv \ddot h / \dot h  = F(h, \dot h)$
 which can be calculated by  eq. (\ref{7.20}).

For a concrete discussion let $ \l \approx 10^2 l^4_{\rm Pl} $ 
 and $m_0 = 10^{-5}m _{\rm Pl} $.  Then both 
conditions  $\b \ll - \sqrt \l$  and $\vert  R_- \vert < l^{-2}_{\rm Pl}$
 are fulfilled.  Now the qualitative behaviour of  the solutions can be summarized: There 
exist two special solutions which approximate the 
ground state $R_-$ for $t \to - \infty$. All other 
solutions have a past singularity $h \to \infty$.  
Two other special solutions approximate 
the ground state $R_-$ for $t \to   + \infty$. 
Further solutions have a future  singularity $h \to \infty$, and all other solutions 
have a power-like behaviour for $t \to \infty$,
 $ a(t)$  oscillates around the  classical dust model $a(t) \sim t^{2/3}$. 
But if we restrict the initial conditions to lie in 
a small neighbourhood of the unstable ground 
state $R_-$, only one of the following three cases appears:

\noindent 
1.   Immediately one goes with increasing values $h$ to a singularity.

\noindent 
2.  As a special case:  one goes back to the de Sitter stage $R_-$.

\noindent
3.  The only interesting one: One starts with a finite 
$l_{\rm Pl}$-valued inflationary era, goes 
over to a GUT-valued second inflation and ends with a power-like 
Friedmann  behaviour.

In the last case to be considered here, let $\l = 0$, $\L > 0$ and $\b < 0$. 
The analogue to  eq. (\ref{7.20})   then reads
\be
0    = h^2/2 - \L /6 + (2h \ddot h - \dot  h^2 +
 6h^2 \dot h)/2m_0^2 \,  .
\ee
Here, always $h \ne 0$  holds, we consider 
only expanding solutions $h > 0$. For $\dot h = 0$ we have
\be
\ddot h =    (\L  m_0^2/3 -m_0^2 h^2)/2h \, .
\ee
For $\ddot h = 0$ we have $\dot h >  m_0^2/6$ and

\be
h = (\L /3 + \dot  h^2/m_0^2)^{1/2}  (1 + 6 \dot h/m_0^2)^{-1/2} \, .
\ee
We obtain the following result:
All solutions approach the de Sitter phase $h^2 = \L /3$
 as $t \to \infty$.   There exists one special solution  approaching 
$ \dot h = -m_0^2/6$ for $ h \to \infty $,
 and all solutions have a past singularity $h \to \infty$.
 For a sufficiently small value $\L$
 we have again two different inflationary eras  in most of all models.

\subsection{The generalized Bicknell theorem}\label{s75}

In this section we derive a general equivalence theorem between a non-linear 
Lagrangian $ L(R)$  and a minimally coupled  scalar field $\phi$ with a general 
potential with Einstein's theory. Instead of $\phi$  we take
\be\label{7.24f}
\psi = \sqrt{2/3} \  \phi\, .
\ee
This is done to avoid square roots in the exponents. 
Then the Lagrangian for the scalar field reads
\be\label{7.23}
 \tilde   L = \tilde R/2 - 3 \tilde g^{ij} \psi_{;i} \psi_{;j}/4 
 + V(\psi) \, .
\ee
At ground states $\psi = \psi_0$, defined by
 $\pa V/ \pa \psi = 0$  the effective mass is
\be\label{7.24}
          m= \sqrt{2/3}  \sqrt{ \pa^2 V/ \pa \psi^2   }     \,  ,   
\ee
cf.  eqs. (\ref{7.3}) and (\ref{7.24f}). The variation $ 0 = \d \tilde L/ \d \psi$ gives
\be\label{7.25} 
         0 =\pa V/ \pa \psi + 3 \, \tilde g^{ij} \tilde \nabla_i \tilde  \nabla_j
\,   \psi / 2
\ee
and Einstein's equation is
\be\label{7.26}
\tilde E_{ij} = \kappa \tilde T_{ij}
\ee
with
\be\label{7.27}
\kappa \tilde T_{ij} =  3 \psi_{;i}\psi_{;j} /2 +
   \tilde g_{ij} \left(  V(\psi) - \frac{3}{4} \tilde g^{ab} \psi_{;a}\psi_{;b}
\right)        \, .
\ee
Now, let
\be\label{7.28}
   \tilde g_{ij} =e^\psi  g_{ij} \, .
\ee
The conformal transformation  eq. (\ref{7.28}) shall 
be inserted into eqs.  (\ref{7.25}),  (\ref{7.26}) and  eq. (\ref{7.27}).
 One obtains from  eq. (\ref{7.25})  with 
\bea\label{7.29}
\psi^{;k} := g^{ik} \psi_{;i}       \nonumber \\
\Box \psi + \psi^{;k}\psi_{;k} = - 2 (e^\psi \pa V/\pa \psi )/3
\eea
and from   eqs. (\ref{7.26}),  (\ref{7.27})
\be\label{7.30}
E_{ij} = \psi_{; ij} + \psi_{;i}\psi_{;j} +g_{ij} 
\left(  e^\psi V(\psi) - \Box \psi   - \psi_{;a} \psi^{;a}  \right) \, .
\ee
Its trace reads
\be\label{7.31}
        -R = 4 e^\psi  V(\psi) - 3 \Box \psi   - 3 \psi_{;a} \psi^{;a} \, .
\ee
Comparing with  eq. (\ref{7.29}) one obtains
\be\label{7.32}
        R = R(\psi) =   - 2e^{-\psi} \pa \left( e^{2\psi} V(\psi) \right) /
 \pa \psi \, .
\ee
Now, let us presume $\pa R/\pa \psi  \ne  0$,
 then  eq. (\ref{7.32}) can be inverted as
\be\label{7.33}
 \psi = F(R) \,   . 
\ee 
In the last step,  eq. (\ref{7.33}) shall be inserted into eqs. 
   (\ref{7.29}),  (\ref{7.30}),  (\ref{7.31}).  Because of
\be
F(R)_{; ij}
= \pa F/ \pa R \cdot  R_{; ij} + \pa^2 F/ \pa R^2 \cdot R_{; i}R_{;j}
\ee  
and $\pa F/\pa R  \ne  0$,   eq. (\ref{7.30})   is a fourth-order 
equation for the metric $g_{ ij}$.   We 
try  to find a Lagrangian    $L = L(R)$ such 
that the equation $ \d L \sqrt{-g} / \d g^{ij} =  0$ 
becomes just  eq. (\ref{7.30}). For $L' = \pa L/\pa R  \ne  0$,
  eq. (\ref{7.14})  can be solved to be
\be\label{7.34}
E_{ij} = - g_{ij}R/2 + g_{ij}L/2 L'  - g_{ij}
\Box L'/L'    - L'_{;ij}/L' \, . 
\ee
We compare     the coefficients of the $R_{;ij}$  terms  
in eqs.   (\ref{7.30})  and   (\ref{7.34}), this gives 
\bea
\pa F/\pa R = L''/L' \,  , \qquad {\rm  hence} \nonumber   \\
L(R) = \mu \int_{R_0}^R e^{F(x)} dx + \L_0   \label{7.35}
\eea 
with suitable constants $\L_0$, $\mu$, and $R_0$, $\mu  \ne  0$.
 We fix them as follows: We are interested in a neighbourhood of 
$R = R_0$   and require $L'(R_0) = 1/2$.  Otherwise $L$
 should be multiplied by a constant factor.  Further, a constant translation
of $\psi$  can be used to obtain $F(R_0) = 0$,
 hence $\mu = 1/2$, $L(R_0) = \L_0$, and
\be
L' (R_0)  = \pa F/\pa R(R_0)/2  \ne  0\,  .
\ee
With  eq. (\ref{7.35}) being fulfilled, the traceless 
parts of eqs.   (\ref{7.30}) and   (\ref{7.35})  
 identically coincide. Furthermore, we have
\be
\Box L'/L' = \Box F + F^{;i} F_{;i}
\ee
and it suffices to test the validity of the relation
\be
e^F \,  V(F(R)) = -R/2 + L/2L' \, . 
\ee
It holds
\bea
2L' = e^F\,  , \qquad {\rm i.e.,} \nonumber \\
 e^{2F} V(F(R)) = L - R e^F/2 \, . \label{7.36}
\eea
At $R = R_0$, this relation reads $V(0) = \L_0 - R_0/2$.  
Applying $\pa /\pa R$  to  eq. (\ref{7.36}) gives 
just  eq. (\ref{7.29}), and, by  the way, $V'(0) = R_0/2 -    2\L_0$. In sum,
\be
L(R) = V(0) + R_0/2 +   \int_{R_0}^R e^{F(x)} dx/2 \, ,
\ee
where $F(x)$ is defined via $F(R_0) = 0$,
\be
\psi   = F\left( -2 e^{-\psi}  \pa(e^{2\psi} V(\psi))   / \pa \psi 
 \right) \, .
\ee

Now, let us go the other direction:  Let $L = L(R)$ be given 
such that  at $R = R_0$,  $ L'L''  \ne  0$. By a constant change of   $L$ let 
$L'(R_0) =     1/2$. Define $\L_0 = L(R_0)$, $\psi = F(R) = \ln (2L'(R))$  and 
consider the inverted function $R = F^{-1}(\psi)$.  Then
\be\label{7.37}
V(\psi) = (\L_0 - R_0/2) e^{-2\psi}
- e^{-2\psi} \int_0^\psi   e^x \ F^{-1}(x) dx/2
\ee
is the potential ensuring the  above mentioned conformal equivalence. 
This procedure is possible at all $R$-intervals 
where $L' \, L''  \ne  0$ holds. For analytical 
functions $L(R)$, this inequality can be violated 
for discrete values $R$  only, or one has simply 
a linear function $L(R)$ being Einstein gravity with $\L$-term.

It turned out that this integral eq. (\ref{7.37}) 
 can be evaluated in closed form as follows: 
\be
V(\psi) = L \bigl( F^{-1}(\psi) \bigr) \cdot e^{-2\psi} 
- \frac{1}{2} \,  F^{-1}(\psi)   \cdot e^{-\psi}    \, .
\ee

Examples: \  1. Let $L = \L + R^2$, $R_0 = 1/4$,   then $4R = e^\psi$  and
\be\label{7.38}
V(\psi ) = \L  e^{-2\psi} - 1/16\, .
\ee

2.   Let $L = \L + R/2 + \b R^2 + \l  R^3/12$, \  $R_0=0$, hence 
$\b  \ne  0$ is necessary. We get
\bea
e^\psi -    1 = 4 \b R + \l R^2/2 \qquad {\rm   and} \nonumber \\
V(\psi) = \L e^{-2\psi} +  \nonumber \\ 
 2\b \l^{-1} e^{-2\psi}  \left( e^\psi - 1 - 
16\b^2(3\l)^{-1}
 ((1 + \l(e^\psi - l)/8\b^2)^{3/2} - 1) \right) \, . \label{7.39}
\eea
The limit $\l \to  0$  in  eq. (\ref{7.39})  is  possible and leads to
\be
V(\psi) = \L e^{-2\psi} -  (1 -  e^{-\psi})^2/16\b \, , 
\ee
so we get for $\L =0$ again the potential from   eq. (\ref{7.2}).

Now, let $R_0$ be a non-degenerated ground state, hence
\be
L(R) = \L_0 + (R - R_0)/2 + L''(R_0) (R - R_0)^2/2 +\dots 
\ee
with $L''(R_0)     \ne  0$  and $\L_0 = R_0/4$, 
cf. subsection \ref{t731}. Using  eq. (\ref{7.37})  we get $V'(0) = 0$  and 
\be
V''(0) = R_0/2 - 1 \, / \bigl(  4L''(R_0) \bigr) \, .
\ee

\subsection{The generalized equivalence}\label{t763}

It is worth mentioning that this conformal equivalence 
theorem can be formulated for arbitrary dimensions $n > 2$: Let
\be\label{8.11}
\tilde{\cal L}
 = \tilde  R/2-   \frac{1}{2}\tilde g^{ij} \phi_{\vert i}
 \phi_{\vert j}     + V(\phi)    
\ee
and
\be\label{8.12}
\tilde g_{ij}= e^{\l \phi} g_{ij} \, , \qquad \l = \frac{  2}{ (n-1)(n-2)}
\ee
be the conformally transformed metric. Then 
the solutions of the variation of   eq. (\ref{8.11})  are transformed 
by  eq. (\ref{8.12})  to the solutions of the variation of $ L = L(R)$, where
\be\label{8.13}
 R = - 2e ^{\l \phi} \left(  \frac{nV}{n-2} + \mu \frac{dV}{d \phi}
\right) \, , \qquad  \mu = \sqrt{\frac{n-1}{n-2}}
\ee
is supposed to  be locally, i.e. near  $R = R_0$,   invertible as
\bea
 \phi=F(R) \, ,      \qquad F(R_0)=0 \, , \qquad F'(R_0) \ne 0   \nonumber \\
{\cal L}(R) = \frac{1}{2} R_0 + V(0) + 
 \frac{1}{2} \int_{R_0}^R e^{F(x)/\mu}  dx    \, . \label{8.14}
\eea
The inverse direction is possible  provided 
 $L'(R) \cdot  L''(R) \ne 0$.

\subsection{The fourth order gravity  model}\label{t764}

Now we come to the following  question:   Let $L(R)$  be a $C^3$-function 
fulfilling $L(0) = 0$, $L'(0) \cdot L''(0) < 0$. Then we can write
\be\label{8.15}
{ \cal L} (R) = \frac{R}{2 } + \b R^2 + O(R^3) \,  ,   \qquad \b <0 \,  .  
\ee
We consider the Bianchi-type I vacuum 
solutions which start in a neighbourhood of the  Minkowski spacetime and ask for the 
behaviour as $t   \to \infty $. Applying the
 equivalence theorem we arrive at the
 models discussed before, and this is applicable for
$\vert \, R \,  \vert $  being small enough. The conformal factor 
depends on $t$ only, and therefore, the space of Bianchi-type I models
 will not be leaved, and we can formulate the 
following: In a neighbourhood of Minkowski spacetime, all 
Bianchi-type I  models which represent a stationary point
 of the action eq. (\ref{8.15}), can be integrated 
up to $t   \to \infty $ or  $-  \infty $, let it be $ + \infty $. One singular solution 
is the Kasner solution and all other solutions undergo 
isotropization and have an averaged equation of state $p = 0$ for $t  \to \infty $.

The de Sitter spacetime
\be\label{9.1}
ds^2=dt^2  - e^{2Ht} (dx^2+dy^2+dz^2) \, , \qquad H \ne 0
\ee
is the spacetime being mainly discussed to represent 
the inflationary phase of cosmic evolution. 
However,  a spacetime defined by
\be\label{9.2}
ds^2=dt^2  - \vert t \vert ^{2p} (dx^2+dy^2+dz^2) \, , \qquad p \ne 0
\ee
enjoys increasing interest for these discussions, too. 
Especially, eq. (\ref{9.2}) with $p \ge 1$, $ t > 0$ is called 
power-law inflation; and with $p < 0$, $t < 0$ it is called polar inflation.

Eq.   (\ref{9.2}) defines a self-similar spacetime: if we 
multiply the metric $ds^2$ by an arbitrary positive 
constant $a^2$, then the resulting  $d \hat s^2 = a^2  ds^2$ 
 is isometric to $ds^2$. 
Proof: We perform a coordinate transformation 
$\hat t = at$, $\hat  x = b(a, p) x$ \dots 
On the other hand, the de Sitter spacetime  eq. (\ref{9.1})  is not 
self-similar, because it has a constant non-vanishing curvature scalar.

Power-law inflation is intrinsically time-oriented. 
Proof: The gradient of the curvature scalar defines a temporal 
orientation. \ 
On the other hand, the expanding ($H > 0$) and the contracting ($H < 0$) 
de Sitter spacetime can be transformed into each other by a 
coordinate transformation, because both of them can be transformed 
to the closed Friedmann universe with scale factor $\cosh(Ht)$,
 which is an even function of $t$. \ 
This property  is connected with the fact 
that  eq. (\ref{9.2}) gives a global description, whereas 
 eq. (\ref{9.1}) gives only a  proper subset of the full de Sitter spacetime.

For $p \to \infty$,  eq. (\ref{9.2}) tends to   eq. (\ref{9.1}). Such a statement has 
to be taken with care, even for the case with real functions. Even more 
carefully one has to deal with spacetimes. The most often used 
limit --  the Geroch-limit  of spacetimes -- has the property that a 
symmetry, here we take it to be self-similarity, of all the elements of the sequence must 
also be a symmetry of the limit.

From this it follows that the Geroch limit 
of spacetimes  eq. (\ref{9.2})  with $p \to \infty$  cannot be unique, 
moreover, it is just the one-parameter set   eq. (\ref{9.1})  
 parametrized by  arbitrary values $H > 0$.

\section{Cosmic no hair theorem and Newtonian limit}\label{s96}
\setcounter{equation}{0}

We now discuss the cosmic no hair theorem, which 
tells under  which circumstances the de Sitter spacetime represents an
attractor solution within the set of other nearby solutions. 
  This property ensures the inflationary model to be  a typical solution.  The notion
``cosmic no hair theorem" is chosen because of  its analogous
properties  to the ``no hair theorem" for black holes.

After a general introduction we restrict our consideration to 
spatially flat Friedmann  models. In this section, we choose  
gravitational Lagrangians 
\be
R  \Box \sp k R \sqrt{-g}
\ee
 and linear combinations of them. They are  motivated from trials how to 
overcome the non--renormalizability  [3, 4] of 
 Einstein's theory of gravity. Results are: For arbitrary  $k$, i.e., 
 for arbitrarily large order $2k+4$ of the 
field equation, one can always find examples where the
attractor property  takes place. Such examples necessarily need a non-vanishing
$R\sp 2$-term.  The main formulas  do not depend on the dimension, so one gets
similar results also for 1+1-dimensional gravity and for Kaluza-Klein cosmology.

Over the years, the notion ``no hair conjecture'' drifted to  
``no hair theorem'' without possessing a generally accepted formulation  or even a
complete proof. Several trials have been made to formulate and prove it at
least for certain  special cases. They all have the overall structure: ``For a
geometrically  defined class of spacetimes and physically motivated
properties of the energy-momentum tensor, all the solutions of the gravitational
field equation  asymptotically converge to a space of constant curvature.''

 Weyl wrote: ``The behaviour of every world satisfying certain natural
homogeneity conditions in the large follows the de Sitter solution asymptotically.''
 Barrow and G\"otz   apply the formulation: ``All
ever-expanding universes with $\Lambda >0$ approach the de Sitter spacetime locally.''

Barrow  gave examples that the no hair conjecture fails 
if the  energy condition  is relaxed and pointed out, that
this is necessary to solve the graceful exit problem. He uses the formulation of
the no hair  conjecture ``in the presence of an effective cosmological
constant, stemming  e.g. from viscosity,  the de Sitter spacetime is a stable
asymptotic solution''.

In the three papers [80],  Prigogine et al.  developed a
phenomenological model of particle and  entropy creation.
It allows particle creation from spacetime curvature, but the inverse procedure,
i.e. particle decay into spacetime curvature is forbidden. 
This breaks the  $t\longrightarrow -t$-invariance of the model. Within 
that model, the expanding de Sitter spacetime is an attractor
solution independently of the initial fluctuations; this
means, only the expanding de Sitter solution is thermodynamically possible.

Vilenkin  discussed  future-eternal inflating
universe models:  they must have a singularity if the condition
D: ``There is at least one point $p$ such that for some point $q$ to the future of $p$
the volume of the difference of the pasts of $p$ and $q$ is finite'' is fulfilled.

 In  Maeda [77]  the following argument is given: If the matter 
distribution is too clumpy, then 
a large number of small black holes appears. Then one should
look for an inflationary scenario where these black holes are harmless.
They cannot   clump together to one giant black hole because of the
exponential expansion of the universe; this explains the existing upper bound of the
mass of black holes in the quasi-de Sitter  model: above 
\be
M_{\rm  crit} = \frac{1}{3\sqrt\Lambda}
\ee
there do not exist  horizons; this  restriction is called  cosmic  hoop conjecture.

For the  no hair conjecture for $R\sp 2$ models one uses
 the formulation ``asymptotical
de Sitter, at least on patch''. The restriction ``on patch'' is not strictly
defined but  refers to a kind of local validity of the statement, e.g., in
a region being covered by one single synchronized system of  reference
in which the  spatial curvature is non-positive and the energy conditions
are fulfilled. The Starobinsky model [103] is  one of those which  does not need
an additional  inflaton field to get the desired quasi de Sitter stage, see 
[100, 101, 104] as three of the earliest approaches to this point. 

 A  further result   is that by the addition of a cosmological  term, the 
Starobinsky model leads naturally to double inflation. Let us
comment  this result: It is correct, but one should add that this is
got at the  price of getting a ``graceful exit problem'',  by this phrase
there is meant the problem of how to finish the inflationary phase
dynamically -- in the 
Starobinsky model this problem is automatically solved by the fact that 
the quasi de Sitter phase is a transient attractor only.

The paper  Buchdahl [38] deals with Lagrangians of  arbitrarily high 
order. Its results are applied in [63, 65, 81]  to general  Lagrangians $F(R, \Box)$.
The original idea of
A. D. Sakharov [54] from 1967 
was to define  higher order curvature corrections to the Einstein action to
get a kind  of elasticity of the vacuum. Then the usual
breakdown of  measurements at 
the Planck length,  such a short de Broglie wave length
corresponds to such  a large mass which makes the measuring
 apparatus to a black hole, is softened.

\subsection{Definition  of  an   asymptotic  de Sitter  
 spacetime}\label{t962}

In this subsection we want to compare some possible definitions
of an asymptotic de Sitter spacetime. To this end let us consider the metric
\begin{equation}\label{y3}
ds^2=dt^2 - e^{2\alpha(t)} \sum_{i=1}^n d(x^i)^2
\end{equation}
which is the  spatially flat Friedmann model in $n$
spatial dimensions. We consider all values $n \ge 1$, but then
 concentrate on the usual case $n=3$. The Hubble parameter is
\be 
H= \dot \alpha \equiv \frac{d\alpha}{dt}\, .
\ee
 We get 
\begin{equation}
R_{00}= - n \left( \frac{dH}{dt} + H^2 \right) \, , \qquad
 R = - 2n \dot H   - n (n+1) H^2 \, .
\end{equation}

Then it holds:  The following 4 conditions for metric  (\ref{y3}) are equivalent. 
1: The spacetime  is flat. \ 2: It holds $R=R_{00}=0$. \ 
3: The curvature invariant  $R_{ij} R^{ij} $ vanishes. \ 
4: Either $\alpha = {\rm const.}$ or  
\be 
n=1 \quad {\rm and } \quad \alpha = \ln \vert t  - t_0 \vert + {\rm  const.}
\ee
For the last case with $n=1$ one has to observe that 
\be
ds^2= dt^2 - (t-t_0)^2dx^2
\ee
 represents flat spacetime in
polar coordinates. For the proof we use  the identity
\begin{equation}
R_{ij} R^{ij} = (R_{00})^2 + \frac{1}{n} (R-R_{00})^2 \, .
\end{equation}

An analogous characterization is valid  for the de Sitter
spacetime.  The following 4 conditions for metric   (\ref{y3})  are
equivalent. 1: It is a non--flat spacetime of constant
curvature. \  2: $R_{00}= R/(n+1)= {\rm  const.}  \ne 0$. \ 
3: $(n+1) R_{ij} R^{ij} = R^2 = {\rm  const.}  \ne 0  $. \
 4: Either $H = {\rm  const.}  \ne 0$ or 
\be
n=1  \ {\rm and } \ \Bigl(  
ds^2= dt^2 - \sin^2(\lambda t)dx^2 \ 
{\rm  or} \  ds^2= dt^2 - \sinh^2(\lambda t)dx^2 \Bigr)  \, .
\ee
For $n=1$, the de Sitter spacetime and anti-de Sitter spacetime differ by 
the factor $-1$  in front of the metric only. For $n>1$, 
 only the de Sitter spacetime,  having 
$R<0$, is covered, because the anti-de Sitter spacetime cannot be represented 
as spatially flat Friedmann model. Our result   shows that within the class of
spatially flat Friedmann models, a characterization of the de Sitter 
spacetime using polynomial  curvature  invariants only, is possible.

Next, let us look for isometries leaving the form of the
metric    (\ref{y3})   invariant. Besides spatial isometries, 
the map $\a \to \tilde \a$ defined by 
\begin{equation}
\tilde \alpha(t) = c + \alpha ( \pm t +t_0), \qquad c, \ t_0 \, = \,  {\rm const.}
\end{equation}
 leads to an isometric spacetime. The simplest expressions
being  invariant by such a transformation are $H^2$ and $\dot H$. We
take $\alpha $
as dimensionless, then $H$ is an inverse time and $\dot H$ an
inverse time squared. Let $H \not= 0$ in the following. The expression 
\begin{equation}                                              
\varepsilon := \dot H H^{-2}
\end{equation}
is the simplest dimensionless quantity defined for the 
spatially flat Friedmann models  and being invariant with respect to this
 transformation. Let 
$n>1$ in the following, then it holds: Two metrics of type    (\ref{y3})   are
isometric if and only 
if the corresponding functions $\alpha $ and $\tilde \alpha $
are related  by this transformation. 
It follows: Metric     (\ref{y3})    with $H\not= 0$ represents  the de
Sitter spacetime  if and only if  $\varepsilon \equiv 0$.

All dimensionless invariants containing at
most second order derivatives of the metric can be expressed as 
$f(\varepsilon )$,  where $f$
is any given function. But if one has no restriction to the order, one gets 
a sequence of further invariants
\begin{equation}                                              
\varepsilon _2 = \ddot H H^{-3}, \quad  \ldots, \  \varepsilon _p = 
                 \frac{d^pH}{dt^p} H^{-p-1} \, .
\end{equation}  
 Let $H>0$ in metric (\ref{y3}) with $n>1$. We
call it an asymptotic de Sitter spacetime if  
\begin{equation}
\lim_{t\to \infty} \frac{\alpha(t)}{t} = {\rm const. } \ne 0
\end{equation}
or 
\be
\lim _{t \to \infty}\,  R^2 = {\rm const.} >0 \quad {\rm  and} \quad 
         \lim _{t \to \infty} \, (n+1)R_{ij}R{ij}-R^2=0
\ee
or for some natural number $p$ one has
\be
 \lim_{t \to \infty} \ \varepsilon _j =0 \quad {\rm  for} \quad 1\leq j\leq p \, .
\ee
In general, all these definitions are different. Using the identity
\be 
R_{ij}R^{ij}=n^2(\dot H + H^2)^2 + n (\dot H+nH^2)^2
\ee
 we will see that  all these definitions lead to the same 
result if we restrict ourselves to the set of solutions of the
higher--order  field equations.    The
problem is that none of the above definitions can be
generalized  to inhomogeneous models. One should find a
polynomial curvature  invariant which equals a positive
constant if and only if the  spacetime is locally the de
Sitter spacetime. To our  knowledge, such an invariant
cannot be found in the literature,  but also the
non--existence of such an invariant has not been   proven up
to now.

This situation is quite different for the positive definite 
case: For signature $(++++)$ and 
$S_{ij} = R_{ij}  - \frac{R}{4} g_{ij} $ it holds:
\be
I \equiv  (R-R_0)^2 + C_{ijkl} C^{ijkl}  + S_{ij} S^{ij}  =0
\ee
if and only if  the $V_4$ is a space of constant curvature $R_0$. So 
$I \longrightarrow 0$ is a suitable definition of an
asymptotic  space of constant curvature.
 One possibility exists, however, for the Lorentz signature case, 
 if one allows additional  structure as follows: An ideal
fluid has an energy--momentum  tensor
\be
T_{ij} = ( \rho + p)u_i u_j - p g_{ij} 
\ee
where $u_i$ is a continuous vector field with
$u_i u^i \equiv 1$. For  matter with equation of state $ \rho = - p$, 
the equation  $T^{ij}_{\ \ ;j} \equiv 0$ implies $p=$ const., and so every
solution of  Einstein's theory with such matter is isometric 
to a vacuum  solution of Einstein's theory with a cosmological
term. The  inverse statement, however, is valid only locally:

Given a  vacuum solution of Einstein's theory with a 
$\Lambda $--term, one  has to find  continuous timelike unit
vector fields which need  not to exist from topological
reasons. And if they exist, they  are not at all unique.  So,
it becomes possible to define an  invariant $J$ which vanishes
if and only if the spacetime is de Sitter by  transvecting the
curvature tensor with $u^i u^j$ and/or $g^{ij} $  and suitable
linear and quadratic  combinations of such terms.  Then time
$t$ becomes defined by the streamlines of the vector  $u^i$.
If one defines the asymptotic de Sitter spacetime by 
$J \longrightarrow 0$ as $t \longrightarrow \infty$, then it
turns out, that this definition is   not  independent of
the vector field $u^i$. 

\subsection{Lagrangian $F(R, \Box R, \Box \sp 2R,
\dots ,  \Box \sp k R )$ }\label{t963}

Let us consider the Lagrangian density $L$ given by
\begin{equation}
L = F(R, \Box R, \Box \sp 2R,
\dots  ,  \Box \sp k R) \sqrt{-g}
\end{equation}
where $R$ is the curvature scalar,  $\Box $ the D'Alembertian
and  $g_{ij}$  the metric of a  Pseudo-Riemannian $V_D$ of 
dimension   $D\ge   2$  and  arbitrary  signature;  
 $g  =  - \vert  \det  \, g_{ij}  \vert  $. 
 The main application will be $D=4$ and  metric
signature $(+---)$.  $F$ is supposed to be  a  sufficiently  smooth 
function of its arguments, preferably a polynomial.
 Buchdahl  already  dealt with such kind of  Lagrangians
 in 1951,  but then it became quiet of them for decades.

The  variational  derivative of $L$ with respect  to  the  metric yields the tensor 
\begin{equation}
P\sp{ij}  \  =  - \frac{1}{\sqrt{-g}} \ 
\frac{\delta  L  }{\delta g_{ij} }
\end{equation}
The components of this tensor  read
\begin{equation}
P_{ij} \ = \ G R_{ij} \ - \ \frac{1}{2} F g_{ij} \ - \ G_{;ij}
 \ + \ g_{ij} \Box G \  + \ X_{ij}
\end{equation}
where the semicolon denotes the covariant derivative,
 $R_{ij}$  the Ricci tensor, and
\begin{equation}
 X_{ij} \ = \ \sum_{A=1}\sp k \ \frac{1}{2} g_{ij}
[F_A(\Box\sp{A-1}  R)\sp{;m}  ]_{;m} \  - \  F_{A(;i}[\Box\sp{A-1} 
R]_{;j)}
\end{equation}
having  the round symmetrization brackets in its last  term. 
For $k=0$, i.e. $F = F(R)$, a case considered in subsection \ref{t964}, the
tensor $  X_{ij}$  identically  vanishes.   It  remains  to 
define  the expressions $F_A$, $A=0, \dots,k$ .
\begin{equation}
F_k \ = \ \frac{\partial F}{\partial \Box \sp k R }
\end{equation}
and for $A=k-1, \dots,0$
\begin{equation}
F_A \ = \ \Box F_{A+1} \ + \ 
 \frac{\partial F}{\partial \Box \sp A R }
\end{equation}
and finally $G \ = \ F_0$. 
The brackets are essential, for any scalar $\Phi $ it holds 
\begin{equation}
\Box(\Phi _{;i}) \ - \ (\Box \Phi)_{;i}  \ = \ R_i \sp { \ j}
\ \Phi_{;j}
\end{equation}
Inserting  $\Phi  =  \Box\sp m R$ into this  equation,  one 
gets  identities to be applied in the sequel without further notice.
 The covariant form of energy-momentum conservation reads 
\begin{equation}
P\sp i _{\ j;i} \ \equiv \ 0
\end{equation}
and $P_{ij}$ identically vanishes  if and only 
if $F$ is a divergence, i.e., locally there can be found a
vector  $v\sp i$ such that $F \ = \ v\sp i _{\ ;i}$ holds.  Remark:
Even for compact manifolds without boundary the restriction 
``locally''  is unavoidable, for  example, let  $D=2$ and  $V_2$ be the
Riemannian  two-sphere $S\sp 2$ with arbitrary positive definite metric. 
$R$ is  a divergence,  but there do not exist  continuous
 vector fields $v\sp i$   fulfilling  $R \ = \ v\sp i _{\ ;i}$  
 on the whole  $S\sp 2$. Example: for $m,n \ \ge \ 0$ it holds
\begin{equation}
\Box \sp m  R \ \Box \sp n R \ - \   R \ \Box \sp{m+n} R \ = \
 {\rm  divergence} \, .
\end{equation}
So, the terms $\Box \sp m  R \ \Box \sp n R $ with naturals
$m$ and $n$ can be restricted to the case $m=0$ without loss
of generality. 

\subsection{No hair theorems for higher-order gravity}\label{t964}

 For $n>1$, the $n+1$-dimensional de Sitter spacetime  is an attractor 
solution for the field equation derived from the Lagrangian
\be
R^{(n+1)/2} \, . 
\ee
 It  is not an attractor solution for the Lagrangian $R \Box^k R$
and $k>0$. There exist  combinations of coefficients $c_i$, such that 
the de Sitter spacetime is an attractor solution for the field 
equation derived from the more general Lagrangian 
\be\label{y4}
L=c_0 \, R^{(n+1)/2} \  + \  \sum\limits_{k=0}^{m} \ 
c_k \,  R \,  \Box ^k  R \, . 
\ee 
 Idea of Proof: Concerning fourth-order gravity this method 
was previously applied e.g. by Barrow  [64].
  The de Sitter spacetime is an exact solution for the field equation,
if and only if $2RG=(n+1)F$. If we make the ansatz 
\be
\dot  \alpha(t)=1+\beta(t)
\ee 
we get the linearized  field equation $0=\ddot \beta+n\dot \beta$ for the Lagrangian
$R^{\frac{n+1}{2}}$. For the Lagrangian $L=R \Box^k R$ we get the linearized field
equation $\Box^k R=(\Box^k R)_{,0}$. For the characteristic
 polynomial we get a recursive formula such that the next order 
is received from the previous one by multiplying with 
$ \, \cdot \,  x \cdot (x+n)$.  We get the roots $x_1=-n-1$, $x_2=-n$  
($k$-fold), $x_3=0$ ($k$-fold)  and $x_4=+1$. Because of 
the last root the de Sitter spacetime is not an attractor
solution.  For the  Lagrangian (\ref{y4})
we get the characteristic polynomial 
\be                                  
P(x)  =  x(x+n)\left[c_0 + \sum_{k=1}^mc_kx^{k-1}(x + n)^{k-1}(x
- 1)(x + n + 1)\right]
\ee
for the linearized field equation. The transformation
$z=x^2+nx+\frac{n^2}{4}$ gives a polynomial $Q(z)$ which 
can be solved explicitly. So one can find those combinations of the 
coefficients such that  the de Sitter spacetime is an attractor
solution.

It turned out that all the variants of the definition of an
asymptotic de Sitter solution given in subsection \ref{t962} lead to
the same class of solutions, i.e., the validity of the
theorems written below does not depend on which of the variants of definition of 
an asymptotic  de Sitter spacetime listed   is applied. 
For the 6th--order case we can summarize as follows:  Let  
\be
L \ = \ R^2 \ + \ c_1 \ R  \  \Box R
\ee
and 
\be
L_{\rm E} \ = \  R \ -  \ \frac{l^2}{6} \ L 
\ee
 with length  $l>0$. Then the following statements are equivalent.

\medskip

1. The Newtonian limit of $L_{\rm E}$ is well--behaved, and the
potential $\phi$ consists of terms 
$ e^{-\alpha r}/r$ with $\alpha \ge 0$ only.

2. The de Sitter spacetime with $H= 1/l $ is an
attractor solution for $L$ in the set of spatially flat
Friedmann models, and this can already be seen from the
linearized field equation.

3. The coefficient $c_1 \ge 0$, and the graceful exit problem 
is solved for the quasi de Sitter phase $H\le 1/l$ of $L_{\rm  E}$.

4. $l^2= l_0^2+l_1^2$ such that  $ l^2 \,  c_1 = l_0^2 \,  l_1^2$ has a
solution with $0\le l_0<l_1$.

5. $0\le c_1 < {l^2}/{4}$.

A formulation which includes also the marginally well-behaved cases 
reads as follows: Let  $L$ and $L_{\rm E}$ as in the previous result, then the 
following statements   are equivalent:

1. The Newtonian limit of $L_{\rm  E}$ is well--behaved, for the
potential $\phi$ we allow ${1}/{r}$ and terms like
\be
\frac{P(r)}{r} e^{-\alpha r} \qquad {\rm  with} \qquad  \alpha > 0
\ee
 and a  polynomial $P$.

2. The de Sitter spacetime with $H= {1}/{l}$ cannot be
ruled out to be an attractor solution for $L$ in the set of
spatially flat Friedmann models if one considers the linearized
field equation only.

3. $L_{\rm  E}$ is tachyonic--free.

4. $l^2= l_0^2+l_1^2$ with  $l^2 \,  c_1 = l_0^2 \,  l_1^2$ has a
solution with $0\le l_0 \le l_1$.

5. For the coefficients we have $0\le c_1 \le \frac{l^2}{4}$.

\subsection{Discussion of no hair theorems}\label{t966}

In subsection \ref{t964} we have shown: The results of the Starobinsky model   
  are structurally stable with respect to
the addition of  a sixth--order term $ \sim   R \Box R$, if the coefficients fufil 
certain inequalities. For the eighth-order case we got:  For  
\be
L=R^2 + c_1 R \Box R + c_2 R \Box \Box R ,  \qquad c_2 \ne
0
\ee  
and the case $n>1$ the de Sitter spacetime with $H=1$ is  an
attractor solution in the set of spatially flat 
Friedmann  models if and only if the following inequalities are 
fulfilled: 
\be
0<c_1< \frac{1}{n+1}, \qquad 0<c_2< \frac{1}{(n+1)^2}
\ee
and
\be
c_1> - (n^2+n+1) c_2 + \sqrt{ (n^4+4n^3+4n^2)c_2^2 + 4 c_2 }
\ee
These inequalities define  an open region in the $c_1-c_2-$plane whose 
boundary contains the origin; and for the other boundary
points  the linearized equation does not suffice to decide the
attractor  property. 

This situation shall be called
``semi--attractor'' for simplicity. In a general context this notion is used to
describe a situation where all Lyapunov coefficients have non-positive real parts and at
least one  of them is purely imaginary.

In contrary to the   6th--order case, here we do not have  a one--to--one
correspondence, but a non--void open intersection with that
parameter set  having the Newtonian limit for $L_E$  well--behaved.

To find out, whether another de Sitter spacetime with an 
arbitrary Hubble parameter $H > 0$ is an attractor solution
for  the eighth--order field equation following from the 
above  Lagrangian, one should remember that $H$ has the
physical  dimension of an inverted time, $c_1$ is a time squared,
$c_2$ is  a time to  power 4. So, we have to replace $c_1$ by
 $c_1 H^2$ and   $c_2$ by $c_2 H^4$ in the above dimensionless inequalities 
to  get  the correct conditions. Example: 
$$
0<c_1 H^2 < \frac{1}{4} \, .
$$

Let us summarize: Here  for a theory 
of gravity of order higher than four the Newtonian limit and
the attractor  property of the de Sitter spacetime are systematically
compared. It should be noted that the details of the theory sensibly
depend  on the numerical values of the corresponding coefficients.
So, no general overall result about this class of theories is
ever to be expected.

We have found out
that for the class of theories considered here, 
one of the typical indicators of instability -- 
cosmological runaway--solutions --  need not to exist, even 
for an arbitrarily high order of the field equation. 
It is an additional satisfactory result that  
this takes place in the same range of parameters where the Newtonian limit is well behaved.

\subsection{The Newtonian limit of 4th-order  gravity}
Let  us consider the gravitational theory defined by the Lagrangian
\be\label{12.1}
 L_{\rm g}=  (8\pi G)^{-1} \Bigl(
R/2 + (\a R_{ij}R^{ij} + \b R^2) l^2 \Bigr) \, . 
\ee
$G$ is Newton's constant, $l$ a coupling length and $\a$
 and $\b$  numerical parameters. $R_{ij}$  and $R$  
are the Ricci tensor and its trace. Introducing the 
matter Lagrangian $L_{\rm m}$  and varying  $  L_{\rm g}
   +  L_{\rm m}$   one obtains the field equation
\be\label{12.2}
     E_{ij}  + \a H_{ij} + \b G_{ij}  = 8\pi G \,  T_{ij}  \,   .         
\ee
For $\a = \b = 0$  this reduces to
 General Relativity Theory. The explicit expressions $H_{ij}$ and $ G_{ij}$
 can be found in  Stelle  [59].

In a well-defined sense, the weak-field slow-motion limit 
of Einstein's theory is just Newton's theory. In the following,  
we consider the analogous problem for fourth-order
 gravity eqs.  (\ref{12.1}),  (\ref{12.2}).

The slow-motion limit can be equivalently described 
as the limit $c \to \infty$, where $c$  is the  velocity of  light. 
In this sense we have to  take  the  limit 
$G \to  0$  while $G \cdot  c$  and $l$  remain constants. 
Then the energy-momentum tensor $T_{ij}$ 
 reduces to the rest mass density $\rho$:
\be\label{12.3}
     T_{ij}=    \delta^0_i \delta^0_j \rho  \,    ,    
\ee
$x^0=     t $ being the time coordinate. The metric  can be written as
\be\label{12.4}
     ds^2 =  (1 - 2\phi) dt^2
 - (1 + 2\psi) (dx^2 + dy^2 + dz^2)          \, .
\ee
Now eqs.   (\ref{12.3})  and   (\ref{12.4}) will be inserted into  
eq. (\ref{12.2}).  In our approach, products and time derivatives 
 of   $\phi$   and $\psi$ can be neglected,  i.e.,
\be
R=4 \D \psi -2 \D \phi \, ,    \qquad {\rm where} \qquad 
\D f =f_{,xx} + f_{,yy} + f_{,zz} \, .
\ee
Further  $R_{00} = - \D \phi$, $H_{00} = -2 \D R_{00} - \D R$
  and $G_{00} = -4 \D R$, where $l = 1$.

Then it holds: The
 validity of the $00$-component and of the trace of  eq. (\ref{12.2}),
\be\label{12.5}
 R_{00} - R/2 + \a H_{00} + \b G_{00} = 8 \pi  G \rho    
\ee
and
\be\label{12.6} 
- R - 4(\a + 3 \b ) \D R = 8 \pi G \rho \,  ,
\ee
imply the validity of the full  eq. (\ref{12.2}).

Now, let us discuss eqs. (\ref{12.5}) and  (\ref{12.6})
 in more details: Eq.   (\ref{12.5})  reads
\be\label{12.7}
- \D \phi    - R/2 +\a (2 \D \D \phi - \D R) - 4 \b \D R = 8 \pi G \rho \, .  
\ee
Subtracting one half of  eq. (\ref{12.6})  yields
\be\label{12.8}
- \D \phi   + 2 \a  \D \D \phi + (\a + 2 \b) \D R = 4 \pi G \rho \, .  
\ee
For $\a  + 2 \b  = 0$  one obtains
\be\label{12.9}
- (1 -  2 \a  \D ) \D \phi   = 4 \pi G \rho 
\ee
and then   $\psi  = \phi$  is a solution of eqs.  (\ref{12.5}),  (\ref{12.6}). 
For all other cases the equations for 
$\phi$  and $\psi $ do not decouple immediately, but, 
to get equations comparable with Poisson's equation 
we apply $\D$ to   eq. (\ref{12.6})   and continue as follows.

     For $\a + 3\b  = 0$  one gets from  eq. (\ref{12.8}) 
\be\label{12.10}
- (1 -  2 \a  \D ) \D \phi   = 4 \pi G ( 1 + 2 \a \D /3 )  \rho   \, .
\ee
The $\D$-operator applied to the source term 
in  eq. (\ref{12.10})   is only due to  the application  of 
$\D$ to the trace, the original equations
   (\ref{12.5}),  (\ref{12.6})  contain only $\rho$ itself.

For $\a = 0$ one obtains similarly the equation
\be\label{12.11}
- (1 + 12 \b  \D ) \D \phi   = 4 \pi G ( 1 + 16 \b \D )  \rho   \, .
\ee

For all other cases - just the cases not yet covered 
by the literature - the elimination of $\psi$ from  the 
system   (\ref{12.5}),  (\ref{12.6})  gives rise to a sixth-order equation
\be\label{12.12}
- \bigl(1 + 4(\a + 3 \b)   \D \bigr)   ( 1 - 2 \a \D)
  \D \phi   =   4 \pi G \bigl( 1 + 2(3 \a + 8 \b) \D \bigr)  \rho   \, .
\ee

Fourth order gravity is motivated by quantum-gravity 
considerations and therefore, its long-range behaviour 
should be the same as in Newton's theory. 
Therefore, the signs of the parameters $\a$, $\b$ should 
be chosen to guarantee an exponentially vanishing and 
not an oscillating behaviour of the fourth-order terms:
\be\label{12.13}
\a \ge 0 \, , \qquad   \a + 3 \b \le 0 \, .
\ee
On the other hand, comparing parts of  eq. (\ref{12.12})  with  
the Proca equation it makes sense to  define the masses 
\be\label{12.14}
m_2 = \left(2 \a  \right)^{-1/2} \quad {\rm and} \quad 
m_0 = \left(  -4(\a + 3\b )    \right)^{-1/2} \, .
\ee
Then  eq. (\ref{12.13}) prevents  the masses of the spin 2 and spin 0 gravitons to 
become imaginary.

Now, inserting a delta source $\rho  = m \delta $  into  eq. (\ref{12.12})  one 
obtains for $\phi $ the same result as Stelle  [59], 
\be\label{12.15} 
\phi = m r^{-1} \bigl( 1 + \exp (-m_0 r)/3 - 4 \exp (-m_2 r)/3 \bigr) \, .
\ee
To obtain the metric completely one has also to calculate $\psi$. It reads
\be\label{12.16}
\psi = m r^{-1} \bigl( 1 - \exp (-m_0 r)/3 - 2 \exp (-m_2 r)/3 \bigr) \, .
\ee
For finite values $m_0$ and $m_2$  these are both 
bounded functions,  also for $r \to 0$. In the limits $\a \to  0$, i.e. 
  $m_2 \to \infty$,    and
 $\a + 3\b \to 0$, i.e.  $m_0 \to \infty$,   the terms with   $m_0$ and $m_2$   
in eqs.   (\ref{12.15})  and  (\ref{12.16})
 simply vanish.  For these cases $\phi$  and $\psi $ become
 unbounded as $r \to 0$.

Inserting  eqs. (\ref{12.15}) and   (\ref{12.16}) into the metric  
  (\ref{12.4}),  the behaviour of 
the geodesics shall be studied. First, for an estimation 
of the sign of the gravitational force we take a test 
particle at rest and look whether it starts falling towards 
the centre or not. The result is: for $m_0 \le 2 m_2$,  
gravitation is always attractive,  and for $m_0> 2m_2$  it is attractive 
for large but repelling for small distances.
 The intermediate case  $m_0 = 2m_2$,  i.e., $3\a + 8\b =  0$, is
 already known to  be a special one from  eq. (\ref{12.12}).

Next,  let us study the perihelion advance 
for distorted circle-like orbits. Besides the general 
relativistic perihelion advance, which vanishes 
in the Newtonian limit,  we have an additional 
one of the following behaviour: for $r \to   0$  
and $ r \to \infty $  it vanishes and for  $r \approx 1/m_0$ 
and $r \approx 1/m_2$  it has local maxima,  i.e., resonances.

Finally,  it should be mentioned that the  gravitational field of an extended 
body can be obtained by integrating eqs.   (\ref{12.15}) and   (\ref{12.16}). 
For a spherically symmetric body 
the far field is also of  the type 
\be
 m r^{-1} \Bigl( 1 + a \exp (-m_0 r) + b  \exp (-m_2 r) \Bigr) \, ,
\ee
 and the  factors $a$ and $b$ carry 
information  about the mass distribution inside the body.

\subsection{Higher-order gravity}

One-loop quantum corrections
 to the Einstein equation can be described by curvature-squared terms and 
lead to fourth order
gravitational field equations; their Newtonian limit is described by a 
potential ``Newton + one Yukawa term".
 A Yukawa potential has the form $\exp (-r/l)/r$  and was originally 
used by Yukawa   to describe the meson field.

Higher-loop quantum corrections to the Einstein equation are expected to 
contain terms of the type $R \Box^k R$  in the Lagrangian, which leads to a 
gravitational field equation of order $2k + 4$.
Some preliminary results to this type of equations are already due 
to Buchdahl [38]. For $k=1$, the cosmological consequences of 
the corresponding  sixth-order field equations 
are  discussed in  [66] and  [75].

In the present chapter we deduce the 
Newtonian limit following from this higher order field equation. The Newtonian 
limit of General Relativity Theory is the usual Newtonian theory. 
 From the general structure of the linearized higher-order field equation
 one can expect that for this higher-order 
gravity the far field of the point mass in the Newtonian limit 
is the Newtonian  potential plus a sum of different Yukawa terms. 
And just this form is that one discussed in connection with the fifth force. 
Here we are interested in the details of this connection 
between higher-order gravity and the lengths and coefficients in 
the corresponding Yukawa terms.

Let us start with the Lagrangian
\be\label{12x}
{\cal L} = \left( R + \sum_{k=0}^p \, a_k \, R \Box^k R 
 \right) \cdot \sqrt{-g} 
\, ,  \qquad a_p \ne 0 \, .
\ee
In our considerations we will assume 
that for the gravitational  constant $G$
 and for the speed of  light $c$ 
it holds $G = c = 1$. This only means a special 
choice of units. In eq. (\ref{12x}), $R$ denotes the curvature scalar, 
$\Box$ the D'Alembertian, and $g$ the determinant of the metric. 
Consequently, the coefficient $a_k$ 
 has the dimension ``length to the power $2k + 2$".

The starting point for the deduction of the field 
equation is the principle of minimal 
action. A necessary condition for it is the stationarity of the action: 
\be
- \, \frac{\d {\cal L}}{\d g_{ij}} = 8 \pi \, T^{ij} \,  \sqrt{-g} \, ,
\ee
 where $T^{ij}$ denotes the energy-momentum tensor. The explicit
equations for 
\be 
P^{ij} \,  \sqrt{-g}= - \,  \frac{\d {\cal L}}{\d g_{ij}}
\ee
    are given in [82].  Here we only need the 
linearized field equation. It reads, cf. [75]
\be\label{12y}
P^{ij}  \equiv R^{ij} - \frac{R}{2} \,   g^{ij} + 2 \sum_{k=0}^p \, a_k  [
 g^{ij} \Box^{k+1} \, R - \Box^k  R^{\, ; \, ij} ] = 8 \pi T^{ij}\, , 
\ee
and for the trace it holds:
\be
g_{ij} \cdot P^{ij} =  -  \frac{n-1}{2}  R + 2n   \sum_{k=0}^p \, a_k [
 g^{ij} \Box^{k+1} \, R ]  = 8 \pi T \, .
\ee
$n$ is the number of spatial dimensions; 
the most important application is of cause $n = 3$.
 From now on we put $n = 3$.

\subsection[Newtonian limit in higher-order gravity]{The Newtonian 
limit in higher-order gravity  }

The Newtonian limit is the weak-field static  
approximation. So we use the linearized 
field equation and insert a static metric and an energy-momentum tensor
\be
T_{ij} = \d^0_i \ \, \d^0_j \, \rho \, , \qquad \rho  \ge  0
\ee
into eq. (\ref{12y}).

Without proof we mention that the metric  can be brought into spatially 
conformally flat form,  and so we may  use
\bea 
g_{ij} = \eta_{ij} + f_{ij } \, ,
  \nonumber \\
\eta_{ij} = {\rm  diag} (1, \,  -1, \,  -1, \,  -1)
 \qquad {\rm and}
 \nonumber \\
f_{ij} = {\rm  diag} (-2\Phi, \, -2\Psi , \,  -2\Psi , \,  -2\Psi ) \, .
\eea
 Then the  metric equals
\be\label{12t}
ds^2 = (1 - 2\Phi) dt^2 - (1 + 2\Psi ) (dx^2 + dy^2 + dz^2) \, ,  
\ee
where $\Phi$ and $\Psi$  depend on $x$, $y$ and $z$.
 Linearization means that the metric $g_{ij}$
 has only a small difference to  $\eta_{ij}$; 
quadratic expressions in  $f_{ij}$ and its 
partial derivatives are neglected.  
We especially consider the case of a point mass. In this case it holds: 
$\Phi  = \Phi(r)$, $ \Psi  = \Psi (r)$,  with 
\be
r = \sqrt{x^2+y^2 +z^2} \, , 
\ee
 because of spherical  symmetry and $\rho = m \,  \d$.  
Using these properties, we deduce the field equation
 and discuss the existence  of solutions of the above mentioned type.

At first we make some helpful general considerations: The functions 
$\Phi$ and $\Psi $  are determined
 by eq. (\ref{12y}) for $i = j = 0$ and the trace of eq. (\ref{12y}). If these 
two equations hold, then  all other component-equations are automatically 
satisfied. For the 00-equation we need $R_{00}$:
\be
R_{00} = - \Delta \Phi \, .
\ee
Here, the Laplacian is given  is as usual by 
\be 
\Delta  =  \frac{\pa^2}{\pa x^2} +  \frac{\pa^2}{\pa y^2}
 +  \frac{\pa^2}{\pa z^2} \, . 
\ee
For the inverse metric we get
\be
g^{ij} = {\rm diag} \left( 1/
 (1 - 2\Phi) , \,   - 1/ (1 + 2\Psi ) , \,   - 1/ (1 + 2\Psi ) , \,   - 1/ (1 + 2\Psi )
 \right) 
\ee
and $ 1/  (1 - 2\Phi)  = 1 + 2\Phi  + h(\Phi )$, where $h(\Phi )$ is 
quadratic in $\Phi$  and vanishes  after linearization. So we get 
\be
g^{ij} = \eta^{ij} - f^{ij} \, .
\ee
 In our coordinate system, $f^{ij}$   equals $f_{ij}$  for all $i, j$. 
For the curvature scalar we get
\be
     R = 2(2 \Delta \Psi - \Delta \Phi )     \, .
\ee
Moreover, we need expressions of the type
 $\Box^k \, R$. $\Box R$  is  defined by $ \Box R  = R_{\, ; \, ij} \, g^{ij}$,   
where ``$;$"  denotes the covariant derivative. 
Remarks: Because of linearization we may replace
 the covariant derivative with  the partial one. So we get
\be
 \Box^k \,  R = (~1)^k \, 2( - \Delta^{k+1} \Phi  + 2 \Delta^{k+1} \Psi )    
\ee
and after some calculus
\be\label{12z}
          -8 \pi \rho  = \Delta  \Phi  + \Delta \Psi \,  .    
\ee
We use eq. (\ref{12z}) to eliminate $\Psi$ from the system. 
So we get   an  equation relating $\Phi$   and $\rho  = m \d$.
\be\label{12p}
     -4\pi \left( \rho + 8  \sum_{k=0}^p \, a_k
 (-1)^k  \Delta^{k+1} \,  \rho \right) 
 = \Delta  \Phi  + 6  \sum_{k=0}^p \, a_k  (-1)^k  \Delta^{k+2} \Phi \, .
\ee
In spherical coordinates it holds
\be 
\Delta   \Phi = \frac{2}{r} \Phi_{\, , \, r }  +\Phi_{\, , \, r r } \, ,  
\ee
because $\Phi$  depends on the radial coordinate $r$ only.

We apply the following lemma: In the sense of distributions it holds
\be
\Delta \left( \frac{1}{r} e^{-r/l} \right) = \frac{1}{rl^2}e^{-r/l} - 4 \pi \d \, . 
\ee
Now we are ready to solve the whole problem. We assume
\be
\Phi  = \frac{m}{r} \, \left( 1 +  \sum_{i=0}^q \,
   c_i  \exp (-r/l_i) \right) \, , \quad l_i > 0 \, .
\ee
Without loss of generality we may assume $l_i \ne l_j$
 for $i \ne j$. Then eq. (\ref{12p})  together with that  lemma  yield
\bea
8\pi   \sum_{k=0}^p \, a_k   (-1)^k  \Delta^{k+1} \,  \d =   \sum_{i=0}^q \, 
\left(\frac{c_i}{t_i} + 6 \sum_{k=0}^p \, a_k  (-1)^k   \frac{c_i}{t_i^{k+2}}
\right)  \frac{1}{rl^2}e^{-r/l_i} \nonumber \\  - 4 \pi   \sum_{i=0}^q 
\left(c_i  + 6 \sum_{k=0}^p  a_k   (-1)^k  
\frac{c_i}{t_i^{k+1}} \right) \d   \nonumber \\
+ 24 \,  \pi \sum_{k=0}^p  \sum_{j=k}^p \sum_{i=0}^p
 c_i a_j (-1)^{j+1} \frac{1}{t_i^{j-k}}  \Delta^{k+1} \d 
\eea
where $t_i = l_i^2$ \, ; therefore also $t_i \ne t_j$ for $i\ne  j$.
This equation is equivalent to the system 
\bea
\sum_{i=0}^q c_i = 1/3 \, , \label{12r}
\\
\sum_{i=0}^q \frac{c_i}{t_i^s} = 0\, ,  \qquad s = 1, \dots p \label{12s}
\\
t_i^{p+1} + 6 \sum_{k=0}^p a_k (-1)^k t_i^{p-k }
= 0\, ,  \qquad i=0, \dots q \, . \label{12q}
\eea
From eq. (\ref{12q}) we see that the values 
$t_i$ represent $q + 1$ different solutions of one polynomial. This 
polynominal   has the  degree $p + 1$. Therefore $q \le  p$.

Now we use eqs. (\ref{12r})  and (\ref{12s}) . They can be written in matrix form as
\be
 \left( \begin{array}{c}  1 \dots  1\\  1/ t_0  \dots  1/t_q
 \\ \dots \\  1/ t_0^p  \dots  1/t_q^p  \end{array} \right)  \cdot 
 \left( \begin{array}{c} c_0 \\ c_1 \\ \dots \\ c_q \end{array} 
 \right)  = \left( \begin{array}{c} 1/3 \\ 0 \\ \dots \\ 0 \end{array}  \right)
\ee
Here, the first $q + 1$ rows form a regular matrix, the Vandermonde matrix. 
Therefore, we get
\be
1/t_i^j =  \sum_{k=0}^q    \l_{jk} \, / \,  t_i^k \qquad  j = q + 1, \dots    p
\ee
with certain coefficients $    \l_{jk}$
 i.e., the remaining  rows depend on the first $q + 1$ ones. If $    \l_{j0} \ne 0$
 then the system has no solution. So   $    \l_{j0}= 0$  for all $ q + 1 \le  j \le  p$. 
But for $q < p$ we would get 
\be
1/t_i^q = \sum_{k=1}^q    \l_{q+1\, k} \, / \,  t_i^{k-1} 
\ee
and this is a contradiction to the above stated regularity. 
Therefore $p$ equals $q$. The polynomial in (14) may be written as
\be
6 \cdot  \left( \begin{array}{c} 1 \  1/t_0 \dots  1/t_0^p\\ 
 \dots \\  1 \  1/ t_p  \dots  1/t_p^p  \end{array} \right)  \cdot 
 \left( \begin{array}{c} a_0 \\  \dots \\ (-1)^p a_p \end{array} 
 \right)  = \left( \begin{array}{c}
-t_0 \\ \dots  \\  - \t_p \end{array}  \right)
\ee
This matrix is again a Vandermonde one, i.e., there exists always a unique 
solution $(a_0, \dots  a_p)$, which  are the coefficients of the quantum 
corrections to the Einstein equation,  such that the  Newtonian limit of 
the corresponding gravitational field equation is a sum of Newtonian and 
Yukawa potential with prescribed lengths $l_i$. A more explicit form of the 
solution is given in  section \ref{s126}.

\subsection{Discussion of the weak-field limit}

Let us give some special examples 
of the deduced formulas of the Newtonian 
limit of the theory described by the Lagrangian (\ref{12x}). 
If all the $a_i$ vanish we get of course the usual Newton theory
\be
\Phi = \frac{m}{r} \, , \quad      \Delta \Phi   = -4 \pi \d \, .
\ee
$\Phi $   and $\Psi $ refer to   the metric  according to eq. (\ref{12t}). For
$ p = 0$ we  get for $a_0 <0$
\be
\Phi =  \frac{m}{r} \left[  1 + \frac{1}{3} \, e^{-r/\sqrt{-6 a_0}} \right]
\ee
cf. [59]   and  
\be
\Psi =  \frac{m}{r} \left[  1 - \frac{1}{3} \, e^{-r/\sqrt{-6 a_0}} \right] \, .
\ee
   For $ a_0 > 0$  no Newtonian limit exists.

For $p = 1$, i.e., the theory following from sixth-order gravity
\be
{\cal L} = \left( R + a_0 R^2 +  a_1  R \Box R 
 \right) \sqrt{-g} \, , 
\ee
we get, see  [81]
\be
\Phi = \frac{m}{r} \left[  1 + c_0  e^{-r/l_0} +  c_1  e^{-r/l_1} \right]
\ee
and
\be
\Psi =     \frac{m}{r} \left[  1 - c_0 e^{-r/l_0} - c_1 e^{-r/l_1}  \right]
\ee
where
\be
c_{0,1} = \frac{1}{6} \mp \frac{a_0}{2 \sqrt{9a_0^2 + 6 a_1}}
\ee
and
\be
l_{0,1}= \sqrt{- 3 a_0 \pm \sqrt{9 a_0^2 + 6a_1  }} \, .
\ee
This result is similar in structure but has different coefficients as 
in fourth-order gravity with  included square of the Weyl tensor in the Lagrangian.

The Newtonian limit for the degenerated case $l_0 = l_1$
 can be obtained by a limiting procedure as follows:
 As we already know $a_0 <0$,
 we can choose the length unit such that $a_0 = - 1/3$. 
The limiting case $9 a_0^2 + 6a_1 \to 0$ 
may be expressed by $a_1 = - 1/6 + \e^2$.  
After linearization in $\e$ we get:
\be
l_i = 1 \pm \sqrt{3/2} \, \e \, c_i   = 1/6 \pm 1/(6 \sqrt 6 \e)
\ee
and applying the limit $\e \to 0$ to the corresponding fields $\Phi $ and $\Psi$ we get 
\bea
\Phi  = m/r  \{1 + (1/3 + r/6) e^{-r} \}
\nonumber \\
 \Psi =    m/r  \{1 - (1/3 + r/6) e^{-r} \} \, . 
\eea
For the general case $p > 1$, the potential is a complicated expression, 
but some properties are explicitly known, these hold also for $p = 0, 1$.
 One gets
\be
\Phi  = m/r \, \left( 1 +  \sum_{i=0}^p \,   c_i  \exp (-r/l_i) \right)
\ee
and
\be
\Psi  = m/r \, \left( 1 -  \sum_{i=0}^p \,   c_i  \exp (-r/l_i) \right)
\ee
where $\sum c_i  = 1/3$; $\sum$ means  $\sum_{i=0}^p$  and $l_i$  and $c_i$
 are, up to permutation of indices,  uniquely determined by the Lagrangian.

There exist some inequalities between the coefficients $a_i$, which must be fulfilled in 
order to get a physically acceptable Newtonian limit. By this phrase we mean 
that besides the above conditions, additionally the fields $\Phi$  and
 $\Psi$  vanish for $r \to \infty$  and that the derivatives $d\Phi /dr$ and $d\Psi /dr$
 behave like $O(1/r^2)$. These inequalities express essentially the fact that
the $l_i$  are real, positive, and different from each other. 
The last of these three conditions can be weakened by allowing the
 $c_i$  to be polynomials in $r$  instead of being constants, cf. the example 
with $p = 1$ calculated above.

The equality $\sum c_i = 1/3$  means that the gravitational potential is 
unbounded and behaves, up to a factor 4/3,  like the Newtonian 
potential for $r \approx  0$. The equation $\Phi  + \Psi = 2m/r$
 enables us to rewrite the metric as
\be
ds^2 = (1 - 2 \theta) \left[ (1-    2m/r) dt^2
- (1  + 2m/r)    (dx^2 + dy^2 + dz^2) \right] \, ,
\ee
which is the conformally transformed linearized
 Schwarzschild metric with the conformal factor $1 - 2\theta$, where
\be
\theta  = \frac{m}{r} \sum c_i e^{-r/l_i}
\ee
can be expressed as functional of  the curvature scalar, this is the linearized
 version of  the conformal transformation theorem.
 For an arbitrary matter configuration the gravitational 
potential can be obtained by the usual integration procedure.

\subsection{A homogeneous sphere}\label{s126}
For general $p$ and characteristic lengths $l_i$ fulfilling 
$0 < l_0 < l_1 < \dots < l_p$  we write the Lagrangian as
\bea
{\cal L} =R - \frac{R}{6} \left[
(l_0^2 + \dots + l_p^2) R + (l_0^2 l_1^2 + l_0^2l_2^2
 \dots + l_{p-1}^2 l_p^2) \Box  R + \right.
\nonumber \\
\left. 
(l_0^2 l_1^2l_2^2 + \dots +  l_{p-2}^2 l_{p-1}^2
 l_p^2) \Box^2 R + \dots 
+ l_0^2 \cdot l_1^2 \cdot \dots \cdot l_p^2 \Box^p R \right] 
\eea
the coefficients in front of $\Box^i R$ in this formula read
\be
\sum_{0 \le j_0 < j_1 < \dots < j_i \le p}
 \quad  \prod_{m=0}^i \, l^2_{j_m} \, .
\ee
Using this form of the Lagrangian, the gravitational potential of a 
point mass reads
\bea
\Phi =     \frac{m}{r} \left[   1 + \frac{1}{3}
 \sum_{i=0}^p (-1)^{i+1}   \prod_{j\ne i}
   \vert \frac{l_j^2}{l_i^2} -1 \vert ^{-1} \, 
  e^{-r/l_i} \right] \, ,  \\ \Psi =     \frac{m}{r} \left[
 1 - \frac{1}{3}  \sum_{i=0}^p (-1)^{i+1}  \prod_{j\ne i}
   \vert \frac{l_j^2}{l_i^2} -1 \vert ^{-1} \,   e^{-r/l_i} \right] \, . 
\eea
For a homogeneous sphere of radius $r_0$  and mass $m$  we get
\be 
\Phi =     \frac{m}{r} \left[  1 + \frac{1}{r_0^3}  \sum_{i=0}^p
  e^{-r/l_i} \, l_i^2 \, \tilde c_i  \bigl( 
r_0 \cosh (r_0/l_i) - l_i \sinh (r_0/l_i)  \bigr)  \right] \, , 
\ee
where
\be
\tilde c_i =   (-1)^{i+1}  \prod_{j\ne i}
   \vert \frac{l_j^2}{l_i^2} -1 \vert ^{-1} \, .
\ee

\section*{Acknowledgement}

I thank the Organizers of the 42nd Karpacz Winter School
 for the kind invitation to present these lectures, and I thank several 
colleagues, especially S. Capozziello, M. Francaviglia, A. Mazumdar, 
S. Odintsov,  V. Sahni, A. Starobinsky and L. Urena for useful
 comments to the first version of this paper. 

\section*{Bibliography}

\noindent  [1] K. Lanczos: A remarkable property of the 
Riemann - Christoffel tensor in four dimensions. Annals of 
Math. {\bf  39} (1938) 842-850.

\noindent  [2] H. Weyl: Gravitation und Elektrizit\"at. Sitzungsber. 
Preuss. Akad. d. Wiss. Teil {\bf  1} (1918) 465-480.

\noindent  [3] R. Utiyama, B. de Witt: Renormalization of a classical 
gravitational field interacting with quantized
matter fields, J. Math. Phys. {\bf  3} (1962) 608.

\noindent  [4] K. Stelle: Renormalization of higher-derivative 
quantum gravity. Phys. Rev. {\bf  D 16} (1977) 953-969.

\noindent  [5] H. Weyl: Reine Infinitesimalgeometrie. Mathemat. Zeitschr. 
{\bf  2} (1918) 384-411.

\noindent  [6] H. Weyl: Eine neue Erweiterung der 
Relativit\"atstheorie. Ann. d. Phys. Leipz. (4) {\bf  59} (1919) 101-133.

\noindent  [7] H. Weyl: Elektrizit\"at und Gravitation. 
Physik. Zeitschr. {\bf  21} (1920) 649-650.

\noindent  [8] H.  Weyl:  Physik. Zeitschr. {\bf  22}
 (1921) 473-480.

\noindent  [9] H. Weyl: Electricity and Gravitation. 
Nature {\bf  106} (1921) 800-802.

\noindent  [10] H. Weyl: Raum, Zeit, Materie, (Berlin, 
Springer 1923).

\noindent  [11] E. Schr\"odinger: \"Uber eine bemerkenswerte 
Eigenschaft der Quantenbahnen eines einzelnen Elektrons. 
Zeitschr. f. Physik {\bf  12} (1923) 13-23.

\noindent  [12] H. Weyl: Elektron und Gravitation I. Zeitschr. f. 
Physik {\bf  56} (1929) 330-352.

\noindent  [13] F. London:  Zeitschr. f. Physik {\bf  42} (1927)  375-389.

\noindent  [14] V. Vizgin: Einstein, Hilbert, Weyl: 
Genesis des Programms der einheitlichen 
geometrischen Feldtheorien. NTM-Schriftenr. Leipzig  {\bf  
21} (1984) 23-33.

\noindent  [15] A. Einstein: Addendum to \noindent  [2]; p. 478.

\noindent  [16] P. Dirac: Long range forces and broken symmetries. 
Proc. R. Soc. Lond. {\bf  A 333} (1973) 403-418.

\noindent  [17] E. Reichenb\"acher: Die Eichinvarianz des Wirkungsintegrals 
und die Gestalt der Feldgleichungen in der
Weylschen Theorie. Z. f. Physik {\bf  22} (1924) 157-169.

\noindent  [18] R. Weitzenb\"ock: \"Uber die Wirkungsfunktion in der 
Weylschen Physik 1, 2, 3. Sitzungsber. Akad. d. 
Wiss. Wien, Math.-naturwiss. Kl. 
Abt. IIa, {\bf  129} (1920) 683-696; 697-703; {\bf  130} (1921) 15-23.

\noindent  [19] W. Pauli: Zur Theorie der Gravitation und 
der Elektrizit\"at von Hermann Weyl. Physik. Zeitschr. 
{\bf  20} (1919) 457-467.

\noindent  [20] W. Pauli: Merkurperihelbewegung und 
Strahlenablenkung in Weyls Gravitationstheorie. 
Berichte d. Deutschen Phys. Ges. {\bf  21} (1919) 742-750.

\noindent  [21] F. J\"uttner: Beitr\"age zur Theorie der Materie. Math. 
Annalen {\bf  87} (1922) 270-306.

\noindent  [22] W. Pauli: Relativit\"atstheorie, Enc. math. Wiss. 
Bd. 5, Teil 2, S. 543-775 (Leipzig, Teubner Verl. 1922).

\noindent  [23] R. Bach: Zur Weylschen Relativit\"atstheorie und 
der Weylschen Erweiterung des Kr\"ummungsbegriffs. 
Math. Zeitschr. {\bf  9} (1921) 110-135.

\noindent  [24] A. Einstein: Eine naheliegende Erg\"anzung des 
Fundamentes der allgemeinen Relativit\"atstheorie. 
Sitzungsbericht. Preuss. Akad. d. Wiss. Teil {\bf  1}, (1921) 261-264.

\noindent  [25] E. Reichenb\"acher: Eine neue Erkl\"arung des 
Elektromagnetismus, Z. f. Physik {\bf  13} (1923) 221-240.

\noindent  [26] U. Kakinuma: On the structure of an electron, 
Part I, II. Proc. Phys.-Math. Soc. Japan Ser. {\bf  3}, 
  I  (1928) 235-242; II (1929) 1-11.

\noindent  [27] C. Lanczos: Elektromagnetismus als nat\"urliche 
Eigenschaft der Riemannschen Geometrie. Zeitschr. f. 
Physik {\bf  73} (1932) 147-168.

\noindent  [28] C. Lanczos: Zum Auftreten des Vektorpotentials in der 
Riemannschen Geometrie. Zeitschr. f. Physik {\bf  75} (1932) 63-77.

\noindent  [29] C. Lanczos: Electricity as a natural property of 
Riemannian geometry. Phys. Rev. {\bf  39} (1932)
716-736.

\noindent  [30] C. Lanczos: Ein neuer Aufbau der Weltgeometrie. 
Zeitschr. f. Physik {\bf  96} (1935) 76-106.

\noindent  [31] C. Lanczos: Matter waves and Electricity. Phys. Rev. {\bf  61}
 (1942)  713-720.

\noindent  [32] C. Lanczos: Lagrangian multipliers and Riemannian spaces. 
Rev. Mod. Phys. {\bf  21} (1949) 497-502.

\noindent  [33] C. Lanzcos: Electricity and General Relativity. Rev. 
Mod. Phys. {\bf  29} (1957) 337-350.

\noindent  [34] C. Lanczos: Quadratic action principle of Relativity. 
J. Math. Phys. {\bf  10} (1969) 1057-1065.

\noindent  [35] V. Vizgin: Hermann Weyl, die G\"ottinger 
Tradition der mathematischen Physik und einheitliche 
Feldtheorien. Wiss. Zeitschr. d. 
E.-M.-Arndt-Univ. Greifswald, Math.-naturwiss. Reihe {\bf  33}
  (1984) 57-60.

\noindent  [36] P. Bergmann: Introduction to the theory of relativity
(New York, Prentice Hall 1942).

\noindent  [37] H. Buchdahl: On functionally constant invariants 
of the Riemann tensor. Proc. Cambr. Philos. Soc. {\bf  68}
 (1970) 179-185.

\noindent  [38] H. Buchdahl:  \"Uber die Variationsableitung
von Fundamentalinvarianten beliebig hoher Ordnung. Acta 
Mathematica {\bf  85} (1951) 63-72.

\noindent  [39] H. Buchdahl: On the Hamilton derivatives arising 
from a class of gauge-invariant action principles in a 
$W_n$. J. Lond. Math. Soc. {\bf  26} (1951) 139-149.

\noindent  [40] H. Buchdahl: An identity between the Hamiltonian 
derivatives of certain fundamental invariants in a $W_4$. 
J. Lond.  Math. Soc. {\bf  26} (1951) 150-152.

\noindent  [41] H. Buchdahl: On the gravitational field equations 
arising from the square of the Gaussian curvature. 
Nuovo Cim. {\bf  23} (1962) 141-156.

\noindent  [42] H. Buchdahl: The Hamiltonian derivatives of  a 
class of fundamental invariants. Quart. J. Math. Oxford
{\bf   19}  (1948) 150.

\noindent  [43] H. Buchdahl: A special class of solutions of the 
equations of the gravitational field arising from 
certain gauge-invariant action principles. Proc. Nat. 
Acad. Sci. USA {\bf  34} (1948) 66-68.

\noindent  [44] H. Buchdahl: Reciprocal static metrics and 
non-linear Lagrangians. Tensor {\bf  21} (1970) 340-344.

\noindent  [45] H. Buchdahl: Quadratic Lagrangians and static gravitational 
fields. Proc. Cambr. Philos. Soc. {\bf  74} (1973) 145-148.

\noindent  [46] H. Buchdahl: Non-linear lagrangians and cosmological 
theory. Monthly Not. R. Astron. Soc. {\bf  150} (1970) 1-8.

\noindent  [47] H. Buchdahl: The field equations generated by the 
square of the scalar curvature: solutions of the Kasner type. 
J. Phys. {\bf  A 11} (1978) 871-876.

\noindent  [48] H. Buchdahl: On a set of conform-invariant equations 
of the gravitational field. Proc. Edinburgh Math. Soc. {\bf  10}
  (1953)  16-20.

\noindent  [49] H. Buchdahl: Remark on the equation 
$ \delta   R^2/ \delta g^{ij}  =0$.  Intern. 
J. Theor. Phys. {\bf  17} (1978) 149-151.

\noindent  [50] A. Eddington: Relativity Theory  (in German)  
 (Berlin, Springer 1925).

\noindent  [51] E. Schr\"odinger: Space-time structure (Cambridge University 
Press 1950).

\noindent  [52] C. Gregory: Non-linear invariants and the problem of motion. 
Phys. Rev. {\bf  72}  (1947) 72-75.

\noindent  [53] E. Pechlaner, R. Sexl: On quadratic Lagrangians in 
General Relativity. Commun. Math. Phys. {\bf  2}  (1966) 165-173.

\noindent  [54] A. D. Sakharov: Vakuumnye kvantovye fluktuacii v 
iskrivlennom prostranstve i teoria gravitacii. Dokl. Akad. Nauk 
SSSR {\bf   177} (1967) 70-71; reprinted with comments in
 Gen. Relat. Grav. {\bf  32} (2000) 365.

\noindent  [55] H.-J. Treder:  Ann. Phys. Leipz. {\bf   32} (1975) 
383-400.

\noindent  [56] B. Ivanov: Cosmological solution with string correction. 
Phys. Lett. {\bf   B 198} (1987) 438.

\noindent  [57] D. Hochberg, T. Shimada: Ambiguity in determining 
the effective action for string--corrected Einstein gravity. 
Progr. Theor. Phys. {\bf   78} (1987) 680.

\noindent  [58] V. M\"uller,  H.-J. Schmidt: On Bianchi type I vacuum 
solutions in $R + R^2$ theories of gravitation. I. The isotropic
 case.  Gen. Relat. Grav. {\bf   17} (1985) 769-781.

\noindent  [59] Stelle, K.: Classical gravity with higher derivatives. Gen. 
Relat. Grav. {\bf   9} (1978) 353-371.

\noindent  [60] N. Mio: Experimental test of the law of gravitation at 
small distances. Phys. Rev. {\bf   D 36} (1987) 2321.

\noindent  [61] F. Stacey,  G. Tuck, G. Moore: Quantum Gravity: 
Observational constraints on a pair of Yukawa terms.
Phys. Rev. {\bf   D 36} (1987) 2374.

\noindent  [62] M. Ander,  M. Nieto: Possible resolution of the 
Brookhaven and E\"otv\"os experiments. Phys. Rev. Lett. {\bf   60}  
(1988) 1225.

\noindent  [63]  L. Amendola, A.  Battaglia Mayer, S. Capozziello, 
S. Gottl\"ober, V.  M\"uller,  F. Occhionero, H.-J. Schmidt:
  Class. Quant. Grav. {\bf 10} (1993)  L43.

\noindent  [64] J. Barrow:  Phys. Lett. B {\bf  180} (1986) 335; 
 Phys. Lett. B {\bf  187} (1987) 12.

\noindent  [65] A. Battaglia Mayer, H.-J. Schmidt:  Class. Quant. Grav. {\bf 10}
(1993) 2441.

\noindent  [66] A. Berkin, K. Maeda:  Phys. Lett. B {\bf  245} (1990) 348; 
 Phys. Rev. D  {\bf  44} (1991) 1691.

\noindent  [67] J. Bi\v c\'ak:  Lect. Notes  Phys. {\bf 540} (2000) 1-126; gr-qc/0004016.

\noindent  [68]  G. Bicknell:  J. Phys. A {\bf  7} (1974) 341. 1061.

\noindent  [69] S. Capozziello, G. Lambiase:  Gen. Relat. Grav. {\bf  32}
 (2000) 295, 673.

\noindent  [70] A. Coley, R. Tavakol:  Gen. Relat. Grav.  {\bf 24} (1992) 835.

\noindent  [71] S. Cotsakis, J.  Demaret, Y.  De Rop, L. Querella: 
Phys. Rev. D {\bf  48} (1993) 4595.

\noindent  [72] D. Coule:  Phys. Rev. D {\bf  62} (2000) 124010, gr-qc/0007037; 
 Class. Quant. Grav. {\bf 12} (1995)  455, gr-qc/9408026.

\noindent  [73] G. Ellis: Ann. Rev. Astron. Astrophys. {\bf  22} (1984) 157; 
  J. Math. Phys. {\bf 8} (1967) 1171.

\noindent  [74] S. Gottl\"ober, V.  M\"uller, H.-J.  Schmidt, A. A.  
Starobinsky: Int. J. Mod. Phys.    {\bf D  1} (1992) 257.

\noindent  [75] S. Gottl\"ober, H.-J.  Schmidt, A. A. 
Starobinsky:  Class. Quantum Grav. {\bf  7} (1990) 893.

\noindent  [76] A. Jakubiec, J.  Kijowski:   Phys. Rev. D {\bf  37} (1988) 1406; 
   J. Math. Phys. {\bf 30} (1989) 1073.

\noindent  [77] K. Maeda:  Phys. Rev. D {\bf  37} (1988) 858; 
  Phys. Rev. D {\bf  39} (1989) 3159.

\noindent  [78] V. M\"uller, H.-J. Schmidt:  Gen. Relat. Grav. {\bf  17} (1985)  769;
  Gen. Relat. Grav. {\bf  21} (1989) 489.

\noindent  [79] V. M\"uller, H.-J.  Schmidt, A.  Starobinsky:  Phys. Lett. B {\bf  202}
 (1988)  198.

\noindent  [80] I. Prigogine, J.  Geheniau, E.  Gunzig, P. Nardone:  
Proc. Nat. Acad. Sci. {\bf 85}(1988) 7428; 
 Gen. Relat. Grav. {\bf 21} (1989) 767; 
Int. J. Theor. Phys. {\bf 28} (1989) 927.

\noindent  [81] I. Quandt, H.-J. Schmidt:  Astron. Nachr. {\bf 312} (1991)
 97; gr-qc/0109005.

\noindent  [82]  H.-J. Schmidt: Astron. Nachr. {\bf 303} (1982)   227, gr-qc/0105104; 
  Astron. Nachr. {\bf 303} (1982)  283,  gr-qc/0105105;
 Ann. Phys. (Leipz.) {\bf  41} (1984)  435, gr-qc/0105108; 
 Astron. Nachr. {\bf 306} (1985) 67,   gr-qc/0105107;
  Astron. Nachr. {\bf 307} (1986)  339,  gr-qc/0106037; 
 Astron. Nachr. {\bf 308} (1987)    183, gr-qc/0106035; 
  Class. Quant. Grav. {\bf 5} (1988) 233; 
 Phys. Lett. B {\bf 214}  (1988) 519; 
 Class. Quant. Grav. {\bf 6} (1989) 557; 
Phys. Rev. D {\bf 50} (1994)  5452, gr-qc/0109006;
Phys. Rev. D {\bf  52} (1995) 6198, gr-qc/0106034; 
 Class. Quantum Grav. {\bf  7} (1990) 1023; 
 Phys. Rev. D {\bf 49} (1994) 6354; 
  Phys. Rev. D {\bf 54} (1996)  7906, gr-qc/9404038; 
Grav.  Cosmol. {\bf  3} (1997) 185, gr-qc/9709071;      
  Int. J. Theor. Phys. {\bf  37} (1998) 691, gr-qc/9512007.

\noindent  [83] H.-J. Schmidt, V. M\"uller:  Gen. Relat. Grav. {\bf  17} (1985)
 971.

\noindent  [84] A. A. Starobinsky, H.-J. Schmidt:  Class. Quant. Grav. {\bf  4}
(1987) 695.

\noindent  [85] B. Whitt:  Phys. Lett. B {\bf  145} (1984) 176.

\noindent  [86]  M. Amarzguioui, O. Elgaroy, D.F. Mota, T. Multamaki:
 Cosmological constraints on $f(R)$ gravity theories within the Palatini approach, 
astro-ph/0510519.

\noindent  [87] S. Capozziello, V. F. Cardone, A. Troisi: Reconciling 
dark energy models with $f(R)$ theories, 
 astro-ph/0501426, Phys. Rev.  {\bf  D 71} (2005) 043503.

\noindent  [88] S. Carloni, P. Dunsby, S. Capozziello, A. Troisi: 
Cosmological dynamics of  $R^n$ gravity, gr-qc/0410046, 
  Class. Quant. Grav. {\bf  22} (2005) 4839.

\noindent  [89] T. Clifton, J. Barrow: The existence of G\"odel, Einstein
  and de Sitter Universes, Phys. Rev. {\bf  D 72}
 (2005) 123003, gr-qc/0511076.

\noindent  [90] T. Clifton, J. Barrow: The power of General Relativity,
Phys. Rev.  {\bf   D 72} (2005) 103005.

\noindent  [91] D. Coule: Quantum Cosmological Models,
gr-qc/0412026, Class. Quant. Grav. {\bf   22} (2005) R125-R166.

\noindent  [92] A. Rendall: Intermediate inflation and the slow-roll 
approximation, Class. Quant. Grav. {\bf  22} (2005) 1655.

\noindent  [93] A.  Sanyal,  B. Modak, C. Rubano, E. Piedipalumbo: Noether 
     symmetry in the higher order gravity theory,
     Gen. Relat. Grav. {\bf  37} (2005) 407.

\noindent  [94]  A.  Sanyal: Hamiltonian 
formulation of curvature squared action,
     Gen. Relat. Grav. {\bf 37} (2005) 1957.

\noindent  [95]  F. Schunck, F. Kusmartsev and E. Mielke: 
     Dark matter problem and effective curvature Lagrangians, 
      Gen. Relat. Grav. {\bf 37} (2005) 1427.

\noindent  [96] G. Allemandi, M. Capone, S.  Capozziello:  
 Conformal aspects of Palatini approach in Extended Theories of Gravity, 
Gen. Relat. Grav.  {\bf 38}  (2006) 33.

\noindent  [97] J. Barrow, T. Clifton:  Exact cosmological solutions of 
scale-invariant gravity theories, Class. Quantum Grav. {\bf 23}
 (2006) L1, gr-qc/0509085.

\noindent  [98] J. Barrow, S. Hervik: Anisotropically Inflating Universes,
Phys. Rev. {\bf D 73 } (2006) 023007, gr-qc/0511127. 

\noindent [99]  K. A. Bronnikov, M. S. Chernakova: Generalized 
theories of gravity and conformal continuations,  Grav. Cosmol. 
{\bf  11} (2005) 305;  gr-qc/0601123.

\noindent [100]  V. Sahni, L. A. Kofman: Some self-consistent solutions 
 of the Einstein equations with one-loop quantum gravitational corrections, 
 Phys. Lett.  {\bf A 117} (1986) 275-278.

\noindent [101] I.  Moss and V. Sahni: Anisotropy in the chaotic 
inflationary universe, Phys. Lett.  {\bf  B  178} (1986) 159-162

\noindent [102] Octavio Obregon, L. Arturo Urena-Lopez, Franz E. Schunck: 
  Oscillatons formed by non linear gravity,  Phys. Rev.  {\bf  D  72} (2005) 024004; 
gr-qc/0404012.

\noindent [103]  A. A. Starobinsky: A new type of isotropic cosmological model 
without singularity, Phys. Lett.  {\bf  B  91} (1980) 99.

\noindent [104]  A. A. Starobinsky: Isotropization of
arbitrary cosmological expansion given by  an effective cosmological
constant, Sov. Phys. JETP Lett. {\bf  37}  (1983) 66.

\noindent [105]  T. Biswas, A. Mazumdar, W. Siegel: 
Bouncing Universes in String-inspired Gravity, hep-th/0508194 (2005).

\noindent [106]  S. Nojiri, S. D. Odintsov: Introduction to Modified Gravity 
and Gravitational Alternative for Dark Energy, hep-th/0601213, v.3 (2006).

\noindent [107]  G. Cognola, E. Elizalde, S. Nojiri, S. D. Odintsov, S. Zerbini:
Dark energy in modified Gauss-Bonnet gravity: late-time acceleration and the 
hierarchy problem, hep-th/0601008 (2006).

\noindent [108]  G. Allemandi, M. Capone, S.  Capozziello:  
 Conformal aspects of Palatini approach in Extended Theories of Gravity, 
Gen.  Rel.  Grav.  {\bf  38}  (2006) 33.

\noindent [109]  S. Capozziello, V. F. Cardone, E. Piedipalumbo, C. Rubano:
 Dark energy exponential potential models as curvature quintessence,
   Class. Quant. Grav.  {\bf  23}  (2006) 1205-1216; astro-ph/0507438.
 
\end{document}